\documentclass{IEEEtran}
\usepackage{cite}
\usepackage{amsmath,amssymb,amsfonts}
\usepackage{algorithmic}
\usepackage{graphicx}
\usepackage{textcomp}
\usepackage{cuted}
\usepackage{float}
\usepackage{multicol}
\usepackage{xcolor}
\usepackage{soul}
\usepackage[normalem]{ulem}

\def\BibTeX{{\rm B\kern-.05em{\sc i\kern-.025em b}\kern-.08em
    T\kern-.1667em\lower.7ex\hbox{E}\kern-.125emX}}
\begin{document}
\title{Poynting Vector Spin in Gyromagnetic Medium and its Impact on Backward Power Flow in Waveguiding Structures}
\author{Rajarshi Sen, \IEEEmembership{Graduate Student Member, IEEE}, Sarang Pendharker \IEEEmembership{Member, IEEE}
\thanks{The authors are with the Electronics and Electrical Communication Engineering department, Indian Institute of Technology Kharagpur, Kharagpur, India - 721302 (e-mail: sarang@ece.iitkgp.ac.in).}}

\maketitle

\begin{abstract}
This paper investigates the reactive power and spin of instantaneous Poynting vector in the bulk of gyromagnetic medium. It is shown that the gyromagnetic medium introduces a spin in the Poynting vector. The spin of the instantaneous Poynting vector and the relative strengths of the real and reactive power components are quantified using the Stokes plot method. Using this technique we investigate the presence of backward power propagation in a ferrite-filled waveguide. We present an analytical expression to locate the crossover point separating the forward and backward power propagation, where the real power propagation has a null. We further show that this backward power propagation leads to corresponding opposing surface currents on a waveguide plate of the ferrite-filled waveguide, while the potential difference between the two plates remain symmetric.
\end{abstract}


\section{Introduction}
\label{sec:introduction}

\IEEEPARstart{T}{he} complex time-averaged Poynting vector (CTAPV) is an indicator of power flow and reactive power in a medium. Conventionally, the complex form of CTAPV is of fundamental importance for the determination and extraction of the impedance of microwave networks \cite{JD_jackson_book}. However for propagating waves only the real part of the CTAPV is conventionally considered for investigation as it corresponds to the average power flow. Recently, there has been a renewed interest in the general form of CTAPV, with both its real and imaginary parts gaining importance for investigating the propagation of optical beams\cite{2019Xu}, the near field of Hertzian dipoles\cite{Baumgartner2022}, evanescent waves\cite{2021Nieto,Picardi2019} and optical forces over particles \cite{2019Xu,Antognozzi2016,Nieto-Vesperinas2022}. These investigations involve using correspondence between the CTAPV and the time-varying Instantaneous Poynting Vector (IPV). 

The rotational aspect of IPV has been documented in the pioneering work of Halevi et al. \cite{Halevi:81}, where they demonstrated elliptical trace drawn by the IPV for an inhomogeneous wave. In \cite{2012Litvin}, Litvin demonstrated the relationship between complex CTAPV and elliptical rotation of the IPV in the case of Gaussian and Bessel beams. Litvin further highlighted how the rotational sense of the IPV is directly relevant to the divergence of the Gaussian and Bessel beams. Similarly, Mitri in \cite{Mitri2016} showed the rotation of the Poynting vector around a Poynting vector singularity generated by the superposition of two coherent opposite-handed vortex beams. In \cite{2012Kim} Kim et al. related the reactive part of the complex CTAPV to the spinning nature of the IPV, in surface wave (SW) across a metamaterial-dielectric interface. Specifically, they emphasized the importance of the rotation of the IPV over transverse power transfer across the interface, thus explaining forward and backward power propagation across the dielectric-negative-$\epsilon$ interface. Even though wave propagation in gyrotropic media has been thoroughly investigated, to the extent that it is now a textbook material \cite{dm_pozzar_book,Kong_2008,erogolu_book}, the methodology of using the complex CTAPV that captures the spinning nature of the IPV has not been thoroughly investigated in gyrotropic materials.


At microwave frequencies, gyromagnetic materials have been extensively used to realize isolators and circulators. In spite of gyromagnetic material-based components becoming standard technology, several new designs and approaches have been reported in recent years \cite{Shi_2022,Yan:22,Zhang_2024,Deng2024,He:25,Grachev2024}. 
We see recent contributions to the theory of ferrite circulator designs \cite{2024circulator1,2022circulator3,2023circulator2,2023circulator4,Marzall2021,Olivier2020,Chou2018} and phase sifters \cite{Nafe2015,Kagita2017} and isolators \cite{Ghaffar2019,Noferesti2021}. Ferrite-based filters have also gained recent research interest \cite{2024ferritefilter,2024GaoQ,Popov2023}. In \cite{2023Jemmeli}, authors have proposed a design technique of an ultra-miniaturized ferrite patch antenna, which taps upon the potential of realizing ferrite-based devices with a significantly small footprint. In addition to radiating structures, the interesting scattering effects from cylindrical ferrite rods have also been subject to study \cite{2023zouros}. Ueda et al. in \cite{2022ueda} used the timer-reversal asymmetry of ferrite material in conjunction with the space-inversion asymmetry of a ferrite-based metamaterial transmission to achieve variable phase-shifting nonreciprocity. In addition to these developments, we can also see further interesting works related to gyromagnetic materials in the areas of topological devices \cite{Zhou2020,Xi_2021}, magnonic spin waves\cite{2022Krowne,Tikhonov2019,Tang2021}, and mode-converting waveguides\cite{2019Afshani,2021Afshani}.



The interest in gyrotropy-based devices is also renewed by the enhanced capability to precisely engineer gyrotropic properties with composite materials \cite{Chern:22,Zangeneh-Nejad2020,PhysRevB.109.085429,R_Tuz_2024}. Recently, \cite{Zhang2025} reports that scattering of a Gaussian beam from a gyromagnetic zero-index material introduces a spatiotemporal vortex in the scattered fields. This is experimentally demonstrated at X-band microwave frequencies in substrate waves scattered from YIG based metamaterials. In addition to supporting nonreciprocity\cite{Caloz2018,Abdelrahman2020} gyromagnetic materials also exhibit hyperbolic isofrequency surfaces\cite{Fesenko2019,Tuz2020}, and can also be employed to control photonic spin\cite{2022Sen} and enhance interconversion between microwave and optical fields\cite{Mukhopadhyay2022}. Recently in \cite{2025Sen}, we proposed and experimentally demonstrated photonic spin-dependent absorption in gyromagnetic sub-wavelength rods at permeability-near-zero frequency. All these developments make gyromagnetic material based devices an attractive contender for realizing several miniaturized components ranging from tunable filters, phase shifters, attenuators, to controlled substrate wave propagation at microwave frequencies. In \cite{2022Sen}, we have investigated the effect of gyrotropy on supporting and suppressing different regimes of isofrequency surface topologies, the presence of material-locked photonic spin, and its conflict with structure-induced photonic spin in a ferrite-filled waveguide. However, the understanding of the CTAPV and IPV in gyromagnetic media and in structures with such gyromagnetic media has not been reported. Moreover, a metric to quantify the rotational nature of the instantaneous Poynting vector arising in certain scenarios is missing.

In this paper, we investigate the CTAPV and 
\begin{table}[!h]
\caption{YIG ferrite material specifications}
\label{tab:mat_spec}
\setlength{\tabcolsep}{3pt}
\begin{tabular}{|p{155pt}|p{75pt}|} 
 \hline
 Parameter & Specification \\ 
 \hline
 Magnetic saturation (4$\pi M_s$) & $1820$ Gauss\\
 Magnetic bias ($H_0$) & $3570$ Oe\\
 Lande factor ($g$) & 2.02\\
 Linewidth ($\Delta H$) & $18$\\
 Dielectric permeability ($\epsilon_f$) & $15$\\
 Dielectric loss tangent (tg$\delta$) & $2\times10^{-4}$\\
 \hline
\end{tabular}
\end{table}
rotating IPV for plane wave propagation in the bulk of a gyromagnetic medium. We show that for any arbitrary direction of propagation in such a medium, the imaginary component of CTAPV is tangential to the isofrequency surface, while being normal to the direction of bias. The perpendicular nature of the real and imaginary components of CTAPV results in the spinning of the IPV. To quantify the spin of IPV, we propose a Stokes parameter plot over the Poincar\'e circle. Using these Stokes plots, we graphically represent the spin of the IPV across wave guiding structures with and without gyrotropic material. The Stokes plot analysis of such applications reveals the relation between the gyrotropy-enforced modifications in the spin profile of the IPV and corresponding backward power propagation in the medium.

This paper is organized as follows: In section~\ref{sec:bulkprop}, we investigate the complex CTAPV and time-varying IPV for bulk propagation in a gyromagnetic ferrite. In section~\ref{sec:quantify_complpower}, we propose a simplified mechanism to quantify the spin of the IPV using the Stokes plot representation. We consider a ferrite-filled waveguide in section~\ref{sec:complex_power_waveguide}, and using the Stokes plot analysis, we investigate the backward power propagation phenomenon and corresponding phenomena. In section~\ref{sec:surf_wave_propagation}, we use the Stokes parameter plot to show the role of reactive power in supporting discontinuous power propagation across an air-ferrite interface supporting nonreciprocal SW propagation. The implications of these fundamental observations and the technological possibilities are briefly discussed in section~\ref{sec:disc}. Finally, we conclude in section~\ref{sec:conc}.

\section{Poynting vector in bulk of gyromagnetic media}
\label{sec:bulkprop}

In order to investigate the nature of the Poynting vector in the bulk of gyrotropic media, we select YIG ferrite following the specifications given in Table~\ref{tab:mat_spec}. YIG ferrite is one of the ferrite materials offering the least amount of gyromagnetic loss with $\Delta H=18$ Oe. The standard permeability tensor for a $\hat{z}$-biased gyromagnetic ferrite is given as
\begin{equation}
    \overset{\leftrightarrow}{\mu}_r
    =\begin{bmatrix}
        \mu^\prime&-j\kappa^\prime&0\\
        j\kappa^\prime&\mu^\prime&0\\
        0&0&1\\
    \end{bmatrix}
\end{equation}
where the permeability term $\mu^\prime$ and gyrotropy term $\kappa^\prime$ governs the wave propagation in the medium. This permeability tensor and all the mathematical formulations in this paper correspond to the basis function $e^{j(\vec{k}\cdot\vec{r}-\omega t)}$.  
\subsection{Computation of isofrequency surfaces in bulk propagation.}
\begin{figure}
    \centering
    \includegraphics{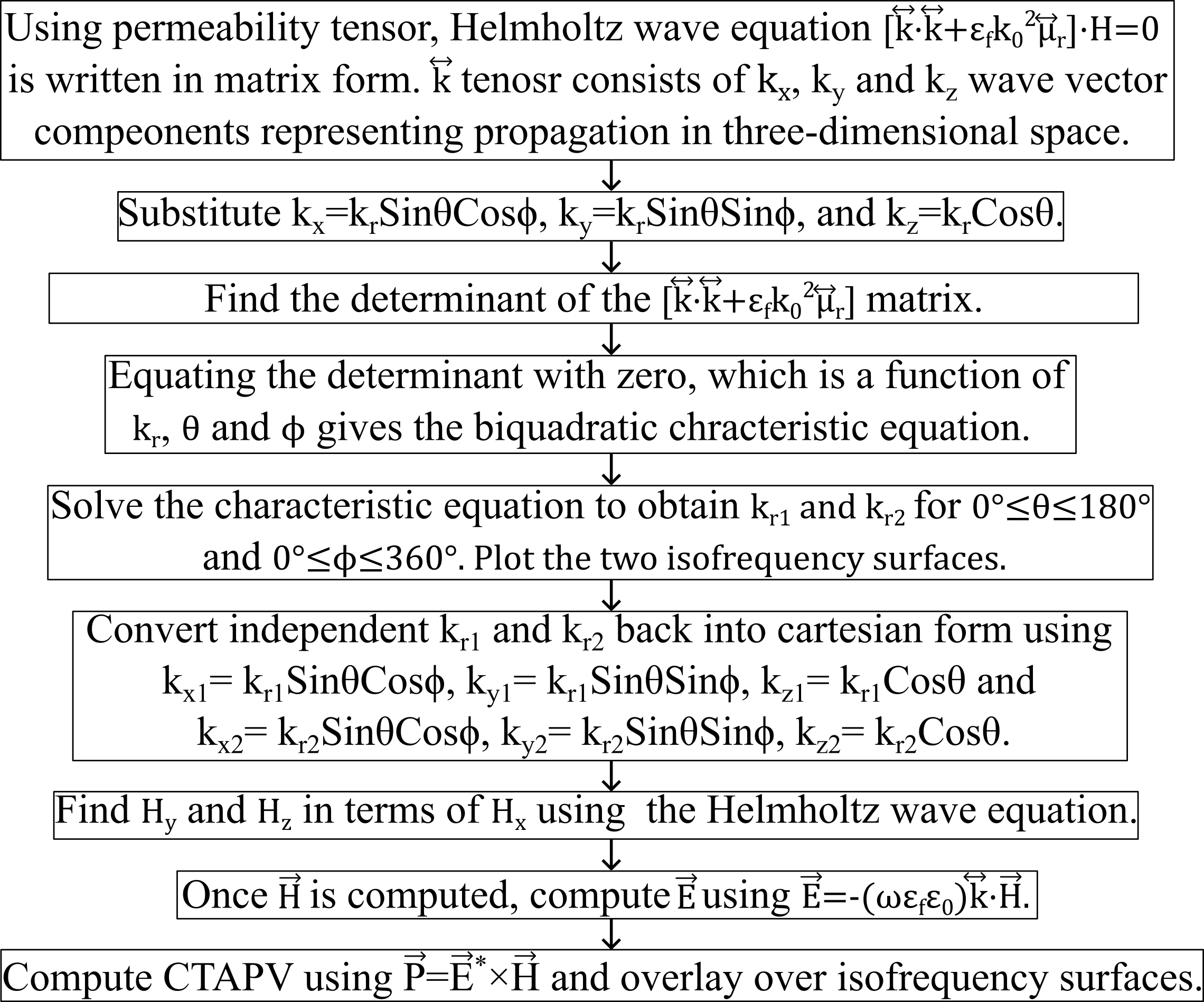}
    \caption{Flow chart describing the steps involved in the computation of isofrequency surfaces and the corresponding Poynting vector.}
    \label{fig:flow_chart_1_bulk}
\end{figure}

Wave propagation in a complex medium can be characterized by the isofrequency surfaces defined by the locus of $(k_x,k_y,k_z)$ for a given frequency $\omega$. A general approach for analysis of such uniaxial media is given in Section 3.3E \& Section3.3F, of \cite{Kong_2008}. We adapt this general procedure to compute the isofrequency surfaces and the corresponding fields as per the steps listed in the flow chart shown in Fig.~\ref{fig:flow_chart_1_bulk}. The $\overset{\leftrightarrow}{k}$ tensor in cartesian coordinate system is give by,

\begin{equation}
    \overset{\leftrightarrow}{k}=
    \begin{bmatrix}
        0&-k_z&k_y\\
        k_z&0&-k_x\\
        -k_z&k_x&0\\
    \end{bmatrix}
\end{equation}

This $\overset{\leftrightarrow}{k}$ containing $k_x$, $k_y$ and $k_z$ components takes into account any arbitrary direction of wave propagation in a 3D space.
Now we can compute the Helmholtz wave equation in matrix form while eliminating $\Vec{E}$ as $[\overset{\leftrightarrow}{k}\cdot\overset{\leftrightarrow}{k}+\epsilon_fk_0^2\overset{\leftrightarrow}{\mu}_r]\cdot\Vec{H}=0$.

\begin{equation}\label{eq:helm_eqn_matrix}
\resizebox{\columnwidth}{!}{%
$\displaystyle
\begin{bmatrix}
\epsilon_f k_0^2 \mu' - k_y^2 - k_z^2 & -j\epsilon_f k_0^2 \kappa' & k_x k_z\\
j\epsilon_f k_0^2 \kappa' + k_x k_y & \epsilon_f k_0^2 \mu' - k_x^2 - k_z^2 & k_y k_z\\
k_x k_z & k_y k_z & \epsilon_f k_0^2 - k_x^2 - k_y^2
\end{bmatrix}
\cdot
\begin{bmatrix} H_x\\ H_y\\ H_z \end{bmatrix}
=0
$%
}
\end{equation}
Where $\epsilon_f$ is the dielectric permittivity of the ferrite medium and $k_0$ is the free space propagation constant. Equating the determinant of this matrix with zero shall give us the characteristic equation as
\begin{multline}
(k_x^2 + k_y^2 + k_z^2)\big(k_z^2 + (k_x^2 + k_y^2)\mu^{\prime}\big)
+ \epsilon_f^2 k_0^4(-\kappa^{\prime 2} + \mu^{\prime 2})
\\+ \epsilon_f k_0^2\big[\kappa^{\prime 2} (k_x^2 + k_y^2)
- \mu^{\prime}\big(2k_z^2 + k_x^2(1+\mu^{\prime}) + k_y^2(1+\mu^{\prime})\big)\big]
= 0
\end{multline}
Here we can substitute $k_x=k_r\sin\theta\cos\phi$, $k_x=k_r\sin\theta\sin\phi$ and $k_z=k_r\cos\theta$ and get
\begin{multline}
0.5 \Big[
k_r^4(1 + \mu^{\prime})
+ 2\epsilon_f^2 k_0^4(-\kappa^{\prime 2} + \mu^{\prime 2})
+ \epsilon_f k_0^2 k_r^2(\kappa^{\prime 2} - \mu^{\prime}(3 + \mu^{\prime}))
\\- k_r^2\big(k_r^2(-1 + \mu^{\prime}) + \epsilon_f k_0^2(\kappa^{\prime 2} + \mu^{\prime} - \mu^{\prime 2})\big)
\cos(2\theta)
\Big]
\end{multline}
Next, we solve this bi-quadratic equation to find $k_r$ as a function of $\theta$ and $\phi$. There will be two distinct solutions for $k_r$, denoting the two different isofrequency surfaces.
\begin{multline}
\label{eqn:kr_eqn_3D}
    k_{r1,2}=(\epsilon_fk_0^2((\mu^\prime+3)\mu^\prime+(\kappa^{\prime2}+\mu^\prime-\mu^{\prime2})\cos2\theta\\-\kappa^{\prime2} \pm(8(\kappa^{\prime2}-\mu^{\prime2})((1+\mu^\prime)+(1-\mu^\prime)\cos2\theta)\\+((\mu^\prime+3)\mu^\prime+(\kappa^{\prime2}+\mu^\prime-\mu^{\prime2})\cos2\theta\\-\kappa^{\prime2})^2)^{0.5})/(2((1+\mu^\prime)+(1-\mu^\prime)\cos2\theta)))^{0.5}
\end{multline}
Equation~(\ref{eqn:kr_eqn_3D}) governs are the governing equations for the 3D isofrequency surfaces (IFSs). However, these expressions are independent of $\phi$. Indicating that the IFSs for gyrotropic ferrite are axially symmetric. This gives us the leverage to understand the 3D wave propagation characteristics by considering only a 2D propagation plane corresponding to any value of $\phi$. For simplicity, we select $\phi=0$, limiting our wave propagation only in the $X-Z$ plane. 
The corresponding $k_x$ and $k_z$ are obtained by converting $k_r$ to cartesian form for the specific values of $\theta$ using
\begin{eqnarray}
    k_{x1}=k_{r1}\sin\theta\\
    k_{z1}=k_{r1}\cos\theta
\end{eqnarray}
\begin{eqnarray}
    k_{x2}=k_{r2}\sin\theta\\
    k_{z2}=k_{r2}\cos\theta
\end{eqnarray}

\subsection{Poynting vector on isofrequency surface.}

Considering wave propagation in $X-Z$ having non-zero $k_x$, $k_z$ and $k_y=0$, the Helmholtz wave equation matrix (as in~(\ref{eq:helm_eqn_matrix})) gets simplified as
$[\overset{\leftrightarrow}{k}\cdot\overset{\leftrightarrow}{k}+\epsilon_fk_0^2\overset{\leftrightarrow}{\mu}_r]\cdot\Vec{H}=0$.

\begin{equation}
\label{eq:helm_eqn_matrix_simpl}
\begin{bmatrix}
\epsilon_f k_0^2 \mu' - k_z^2 & -j\epsilon_f k_0^2 \kappa' & k_x k_z\\
j\epsilon_f k_0^2 \kappa' & \epsilon_f k_0^2 \mu' - k_x^2 - k_z^2 & 0\\
k_x k_z &0 & \epsilon_f k_0^2 - k_x^2
\end{bmatrix}
\cdot
\begin{bmatrix} H_x\\ H_y\\ H_z \end{bmatrix}
=0
\end{equation}
We can use this simplified Helmholtz wave equation matrix to find $H_y$ and $H_z$ with respect to $H_x$. Specifically, we take out following two equations
\begin{equation}
\label{eq:Hx_Hy_rel}
    j\epsilon_fk_0^2\kappa^\prime H_x+(\epsilon_fk_0^2\mu^\prime-k_x^2-k_z^2)H_y=0
\end{equation}
\begin{equation}
\label{eq:Hx_Hz_rel}
    k_xk_z H_x+(\epsilon_fk_0^2-k_x^2)H_z=0
\end{equation}
Taking $H_x$ as a constant $H_0$, we find $H_y$ and $H_z$ in terms of $H_x$ using~(\ref{eq:Hx_Hy_rel}) and (\ref{eq:Hx_Hz_rel}). The corresponding magnetic field vector can thus be written as 
\begin{equation}
    \Vec{H}=H_0\left(\hat{x}+j\frac{\epsilon_fk_0^2\kappa^\prime}{k_x^2+k_z^2-\epsilon_fk_0^2\mu^\prime}\hat{y}+\frac{k_xk_z}{k_x^2-\epsilon_fk_0^2}\hat{z}\right)
\end{equation}
A detailed investigation of the effect of gyrotropy on the isofrequency surfaces and the spin of the magnetic field is presented in \cite{2022Sen}. Using this equation of the magnetic field in phasor form, we can find the corresponding electric field vector in its phasor form using Maxwell's equation $\Vec{E}=(-1/\omega\epsilon_f\epsilon_0)\overset{\leftrightarrow}{k}\cdot\Vec{H}$ as
\begin{multline}
    \Vec{E}=H_0\Bigg(\Bigg. j\frac{k_0^2\kappa^\prime k_z}{\epsilon_0\omega(k_x^2+k_z^2-\epsilon_fk_0^2\mu^\prime)}\hat{x} \\+\frac{k_0^2 k_z}{\epsilon_0\omega (k_x^2-\epsilon_fk_0^2)}\hat{y} +j\frac{k_0^2\kappa^\prime k_x}{\epsilon_0\omega(\epsilon_fk_0^2\mu^\prime-k_x^2-k_z^2)}\hat{z}\Bigg.\Bigg)
\end{multline}
As we now have the electric and magnetic field equations in the phasor form corresponding to bulk propagation in the ferrite medium, we can compute the time-varying instantaneous Poynting vector (IPV) $\Vec{P}_{IPV}=\text{Re}(\vec{E}e^{-j\omega t})\times\text{Re}(\vec{H}e^{-j\omega t})$ and complex time-averaged Poynting vector (CTAPV) $\Vec{P}_{CTAPV}=0.5(\vec{E}^*\times\vec{H})$ as

\begin{strip}
    \begin{multline}
    \label{eq:inst_pv_bulk_yig}
    \vec{P}_\text{IPV}=H_0^2k_0^2
    \Bigg(\Bigg.\Bigg(\Bigg. \frac{k_x(\epsilon_fk_0^2\kappa^{\prime2}(k_x^2-\epsilon_fk_0^2)^2+k_z^2(k_x^2+k_z^2-\epsilon_fk_0^2\mu^\prime)^2)}{2\omega\epsilon_0(k_x^2-\epsilon_fk_0^2)^2(k_x^2+k_z^2-\epsilon_fk_0^2\mu^\prime)^2}\\+\frac{k_x(-\epsilon_fk_0^2\kappa^{\prime2}(k_x^2-\epsilon_fk_0^2)^2+k_z^2(k_x^2+k_z^2-\epsilon_fk_0^2\mu^\prime)^2)}{2\omega\epsilon_0(k_x^2-\epsilon_fk_0^2)^2(k_x^2+k_z^2-\epsilon_fk_0^2\mu^\prime)^2}\cos(2\omega t)\Bigg.\Bigg)\hat{x}-\frac{\kappa^\prime k_x(\epsilon_fk_0^2-k_x^2-k_z^2)}{2\omega\epsilon_0(\epsilon_fk_0^2-k_x^2)(k_x^2+k_z^2-\epsilon_fk_0^2\mu^\prime)}\sin(2\omega t)\hat{y}  \\+  \Bigg(\Bigg.\frac{k_z((k_x^2+k_z^2-\epsilon_fk_0^2\mu^\prime)^2+\epsilon_fk_0^2\kappa^{\prime2}(\epsilon_fk_0^2-k_x^2))}{2\omega\epsilon_0(\epsilon_fk_0^2-k_x^2)(k_x^2+k_z^2-\epsilon_fk_0^2\mu^\prime)^2}+\frac{k_z((k_x^2+k_z^2-\epsilon_fk_0^2\mu^\prime)^2-\epsilon_fk_0^2\kappa^{\prime2}(\epsilon_fk_0^2-k_x^2))}{2\omega\epsilon_0(\epsilon_fk_0^2-k_x^2)(k_x^2+k_z^2-\epsilon_fk_0^2\mu^\prime)^2}\cos(2\omega t)  \Bigg.\Bigg)\hat{z}\Bigg.\Bigg)
    \end{multline}
    \begin{multline}
        \label{eq:pavg_bf}
        \vec{P}_\text{CTAPV}=H_0^2k_0^2\Bigg( \Bigg. \frac{k_x(\epsilon_fk_0^2\kappa^{\prime 2}(k_x^2-\epsilon_fk_0^2)^2+k_z^2(k_x^2+k_z^2-\epsilon_fk_0^2\mu^\prime)^2)}{2\omega\epsilon_0(k_x^2-\epsilon_fk_0^2)^2(k_x^2+k_z^2-\epsilon_fk_0^2\mu^\prime)^2}\hat{x}\\+j\frac{\kappa^\prime k_x(\epsilon_fk_0^2-k_x^2-k_z^2)}{2\omega\epsilon_0(\epsilon_fk_0^2-k_x^2)(k_x^2+k_z^2-\epsilon_fk_0^2\mu^\prime)}\hat{y}+\frac{k_z(\epsilon_fk_0^2\kappa^{\prime 2}(\epsilon_fk_0^2-k_x^2)+(k_x^2+k_z^2-\epsilon_fk_0^2\mu^\prime)^2)}{2\omega\epsilon_0(\epsilon_fk_0^2-k_x^2)(k_x^2+k_z^2-\epsilon_fk_0^2\mu^\prime)^2}\hat{z} \Bigg. \Bigg) 
    \end{multline}
\end{strip}

\begin{figure}
    \centering
    \includegraphics{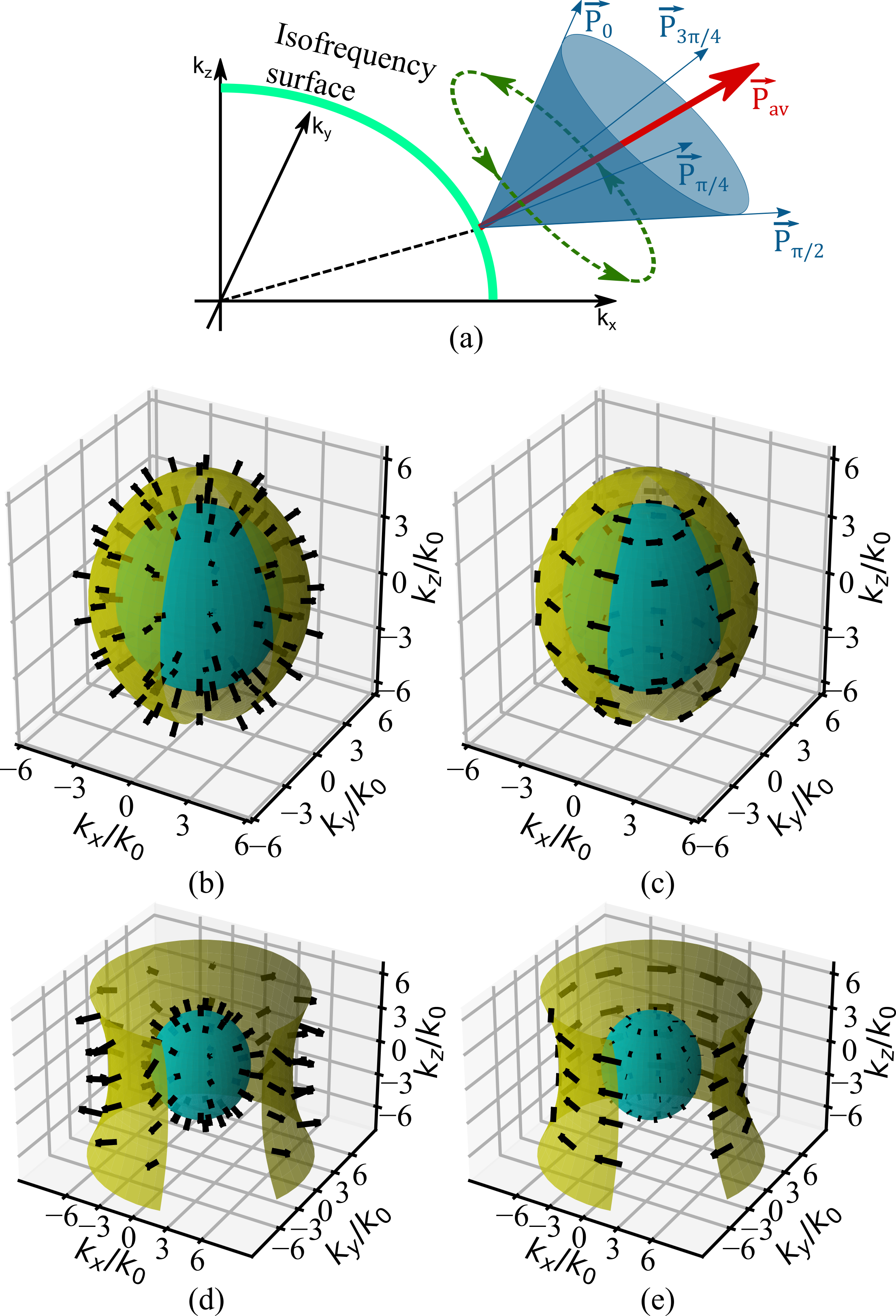}
    \caption{(a) Graphical illustration of the time-varying behavior of the IPV. The real and imaginary components of the CTAPV over the 3D IFSs are shown in (b,c) corresponding to the elliptical regime at $7$ GHz, and (d,e) corresponding to the hyperbolic regime at $11$ GHz. Panels (b) and (d) correspond to the real part of the CTAPV and (c) and (e) correspond to the imaginary part.}    \label{fig:isofreq_surf_pvrot_real_reac_power_quiver}
\end{figure}

Here, $k_0=\omega/c$ is the free space propagation constant, and $c$ denotes the speed of light in vacuum. Interestingly, the IPV expression of~(\ref{eq:inst_pv_bulk_yig}) shows that in addition to the usual in-plane components along $\hat{x}$- and $\hat{z}$-directions, we have an additional gyrotropy-induced time-varying component along the transverse $\hat{y}$-direction. Moreover, we note that the time-varying transverse component of IPV is $90^\circ$ out of phase compared to the in-phase time-varying $\hat{x}$- and $\hat{z}$-components as indicated by their $\sin(2\omega t)$ and $\cos(2\omega t)$ dependencies. The $90^\circ$ phase difference between the time-varying IPV components leads to a rotation of the IPV around the time-invariant component. This temporal rotation of the IPV is schematically illustrated in Fig.~\ref{fig:isofreq_surf_pvrot_real_reac_power_quiver}~(a) by a blue elliptical conical trace around the time-invariant component of the IPV (equivalent to averaging in time) shown using the red arrow. The direction of rotation of the IPV around the time-invariant component is depicted using the green dashed curve having arrows encircling the trace, which is guided by the four time-phases ($\omega t$=$0$, $\pi/4$, $\pi/2$, and $3\pi/4$) of the IPV.

We observe that the movement of the IPV follows an elliptical trace around the real part of CTAPV. This elliptical rotation depends on the additional transverse $\hat{y}$-component, which is usually not observed for a plane wave propagating in the bulk of a conventional non-gyrotropic material along the $X-Z$ plane.

To complete the investigation of power propagation in the bulk of the ferrite material, we look into the CTAPV in the medium. The CTAPV for plane waves in the bulk of the medium (see (\ref{eq:pavg_bf})) has a reactive power component along the transverse $\hat{y}$-axis in addition to the real part along the $X-Z$ plane of propagation. We can observe the correspondence between the CTAPV and IPV as the time-invariant in-plane components of the IPV being equal to the real part of the in-plane components of the CTAPV. Moreover, the out-of-phase time-varying transverse component of IPV equals the transverse reactive power component of CTAPV. This equality leads to the inference that the transverse reactive power component governs the transverse out-of-phase component of the IPV. In other words, the complex CTAPV holds information about the elliptical spin of the IPV around the time-invariant real power propagation. This transverse reactive power is an essential feature of the gyrotropic materials, as can be understood by the direct proportionality of the $\hat{y}$-component with the gyrotropic term $\kappa$. In the absence of gyrotropy, i.e. $\kappa^\prime=0$ will lead to the absence of the transverse reactive component of the complex CTAPV, and IPV instead of following an elliptical spin will follow a linear variation along the $X-Z$ plane. Additionally, reversing the bias orientation will lead to a sign change of $\kappa^\prime$, which will lead to a reversal of the phase of reactive power of the CTAPV and the sense of rotation of the IPV. Thus, the reactive power plays an essential role in the spin of the IPV in the gyrotropic medium. 

Gyromagnetic ferrite can support different isofrequency surface (IFS) topologies based on the combinations of medium parameters $\mu^\prime$ and $\kappa^\prime$ (as we showed in \cite{2022Sen}). Accordingly, we select two different frequencies, $7$ GHz and $11$ GHz, which support the medium parameters necessary for the elliptical and hyperbolic IFS topologies, respectively. Investigating the Poynting vector corresponding to the elliptical and hyperbolic regimes of wave propagation gives us a comprehensive understanding of power propagation behavior in the gyromagnetic ferrite medium. Figure~\ref{fig:isofreq_surf_pvrot_real_reac_power_quiver}(b,c) and (d,e) show the real and reactive components of CTAPV across the IFSs in the elliptical and hyperbolic regimes, respectively. Panels (b) and (d) correspond to the real part of the CTAPV, and the imaginary part for the two regimes is shown in panels (c) and (e). We observe that the real part of CTAPV is normal to the IFS, which is in agreement with previous studies in the literature. However, we also observe the transverse nature of the reactive power component tangential to the surface and transverse to the direction of propagation. This gyrotropy-induced reactive power in the transverse direction is an important property of wave propagation in the gyrotropic medium. It plays a crucial role in governing guided wave propagation involving gyrotropic materials (as we will see in sections~\ref{sec:complex_power_waveguide} and \ref{sec:surf_wave_propagation}). The results shown in Fig.~\ref{fig:isofreq_surf_pvrot_real_reac_power_quiver} corresponding to the CTAPV were computed for a loss-less scenario. However, our investigation, including standard material losses, shows the negligible effect of the losses on the fundamental results reported in this work and are given in the supplementary document.

Conventional guided wave applications often contain a reactive part in the CTAPV. Furthermore, when a gyrotropic material that inherently supports reactive power for bulk modes is introduced to such applications, the combined effect has a significant impact on the power propagation characteristics of the guiding structure. To better understand the role of the reactive power over the overall power propagation, we propose a method of quantifying the spin of the IPV using the complex CTAPV in the next section.

\section{Quantification of Poynting vector rotation using Stokes circles}
\label{sec:quantify_complpower}


\begin{figure}
    \centering
    \includegraphics{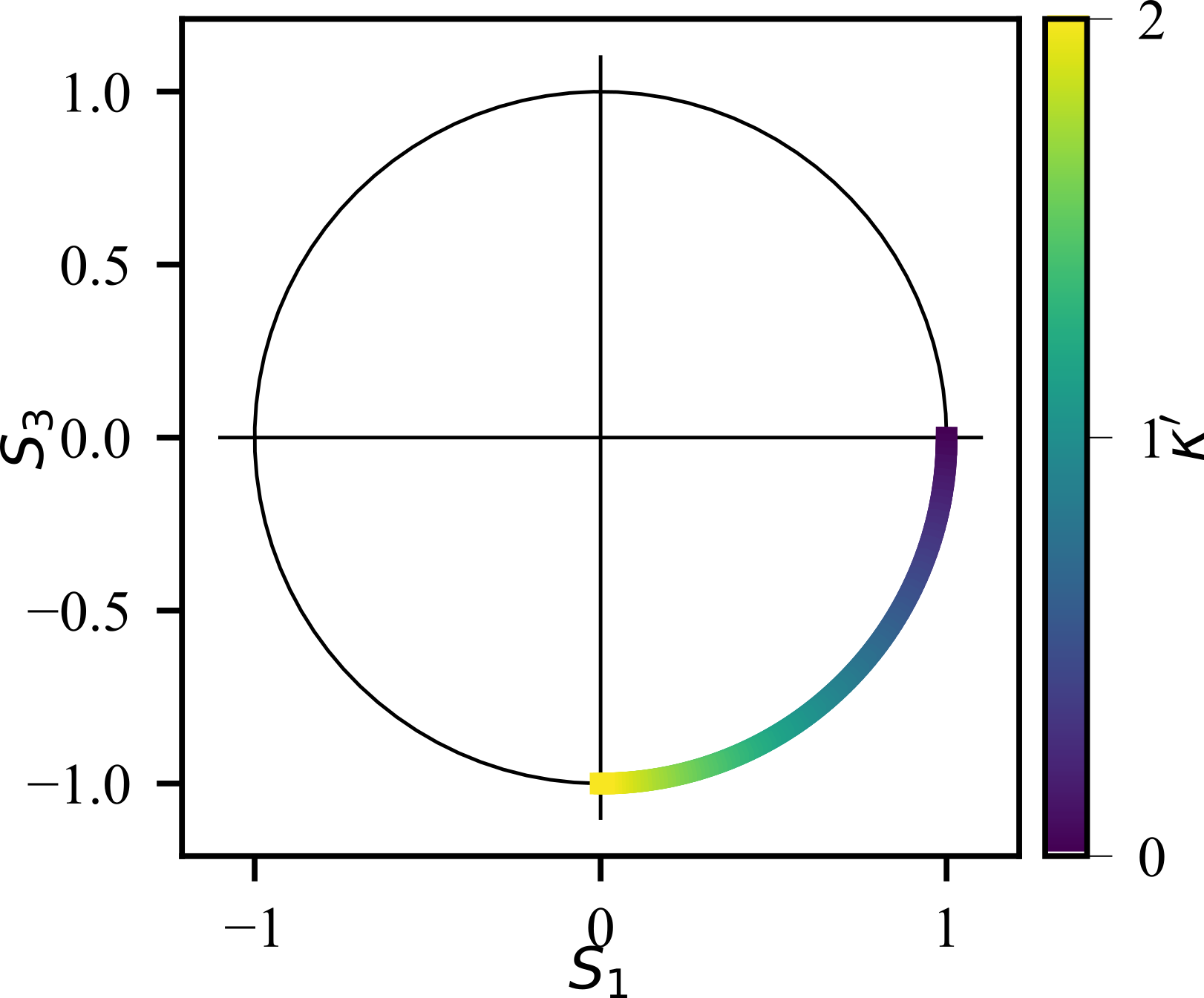}
    \caption{Variation of Stokes parameters $S_1$ and $S_3$ over the Poincar\'e circle corresponding to the variation of $\kappa^\prime$ from $0$ to $2$. We have $\mu^\prime=2$ for this propagation scenario along $\hat{x}$-axis.}
    \label{fig:bulk_TE_stokes_parms_movement}
\end{figure}

Considering wave propagation along $\hat{x}$-axis for a $\hat{z}$-biased ferrite, the medium can support a TEM and a TE mode propagation. The field components corresponding to these two propagation modes are ($E_y$, $H_z$) and ($E_z$, $H_y$, $H_x$), respectively. For the TE bulk mode, we can observe the additional longitudinal field component $H_x$ supported by the medium, which has a $90^\circ$ difference with respect to the transverse component. As a result, the CTAPV has a longitudinal real $P_x=-0.5\text{Re}(E_z^*H_y)$ and a transverse reactive $P_y=0.5\text{Im}(E_z^*H_x$) power component. Using this CTAPV in the complex phasor form, we can quantify the spin of the IPV using the classical approach of computing the spin vector as
\begin{equation}
\vec{S}_p=\text{Im}({\vec{P}_{avg}^*\times\vec{P}_{avg}})    
\end{equation}
However, this spin vector computation is insufficient in capturing the important information about the relative difference in magnitude of the real $|P_x|$ and reactive $|P_y|$ power components. 

In order to quantify both the spin of the IPV as well as the relative difference of magnitude between $|P_x|$ and $|P_y|$, we use the Stokes parameters 
\begin{eqnarray}
S_0&=&|P_x|^2+|P_y|^2 \\ S_1&=&|P_x|^2-|P_y|^2 \\
S_2&=&2|P_x||P_y|\cos\phi \\ S_3&=&2|P_x||P_y|\sin\phi
\end{eqnarray}
Here, $S_0$ corresponds to the magnitude of the CTAPV. The $S_2$ parameter will always be zero for our TE mode of propagation as we have $\phi=90^\circ$. The $S_1$ parameter gives us information about the relative difference between the magnitude of power components. The final parameter $S_3$ is equivalent to the conventional spin vector $\vec{S}_p$ and gives information about the spin of the IPV. Conventionally, these Stokes parameters can be traced over the Poincar\'e sphere. However, since we have $S_2=0$ in this case, we can represent the Stokes parameters over a 2D Poincar\'e circle, which is normalized to the maximum radius of $1$ by dividing $S_1$ and $S_3$ by $S_0$ at every gyrotropy value. 
 
The propagation constant corresponding to the TE mode along $\hat{x}$ can be written as $k_x=k_0\sqrt{\epsilon_f\mu_{eff}}$, where the effective permeability $\mu_{eff}$ is $\mu_{eff}=(\mu^{\prime2}-\kappa^{\prime2})/\mu^\prime$. We consider the elliptical regime for the TE mode of propagation, which takes place when we have $\mu^\prime$ as positive and $|\mu^\prime|>|\kappa^\prime|$. Selecting frequency of $7$ GHz and $\mu^\prime=2$, we vary $\kappa^\prime$ from $0$ to $2$, which corresponds to no-gyrotropy to the maximal gyrotropy that can sustain TE wave propagation in the elliptical regime before forcing $k_x$ to be zero, triggering a gyrotropy-induced cutoff in the medium.

The corresponding Stokes plot normalized with $S_0$ at with respect to variation in the $\kappa^\prime$ is shown in Fig.~\ref{fig:bulk_TE_stokes_parms_movement}, where $S_0$ at each value of $\kappa^\prime$, making the trace move along the unity perimeter of the Poincar\'e circle. We can observe that with increasing value of $\kappa^\prime$ from $0$ to $2$, the Stokes plot moves from the coordinate $S_1=1,S_3=0$ to $S_1=0,S_3=-1$ at which point the $k_x$ becomes $0$, leading to a gyrotropy-induced cutoff. From the point of view of the power components of the CTAPV, these two extreme points refer to the point where there is no reactive power (at the no-gyrotropy condition) and when the magnitude of real power approaches and the magnitude of reactive power near cutoff. Apart from the relative magnitude of the real and reactive power components, the $S_3$ parameter, which increases with the $\kappa^\prime$, indicates the increase in the spin of the IPV, until it becomes circular before the cutoff. Thus, this Stokes plot representation is an effective tool in quantifying both the relative magnitude of the components of the CTAPV as well as the spin of the IPV, both of which are important for gyrotropic materials, especially for waveguiding applications. In the next section, we shall look into the application of Stokes plot analysis corresponding to CTAPV in a ferrite-filled rectangular waveguide.

\section{Gyrotropy controlled Poynting vector in ferrite-filled waveguide}
\label{sec:complex_power_waveguide}

In order to investigate the power propagation behavior in a ferrite-filled rectangular waveguide, it is necessary to first characterize the power propagation scenario in a conventional dielectric-filled waveguide using Stokes plot.

\subsection{Poynting vector spin and Stokes plots in dielectric-filled waveguide}

\begin{figure}
    \centering
    \includegraphics{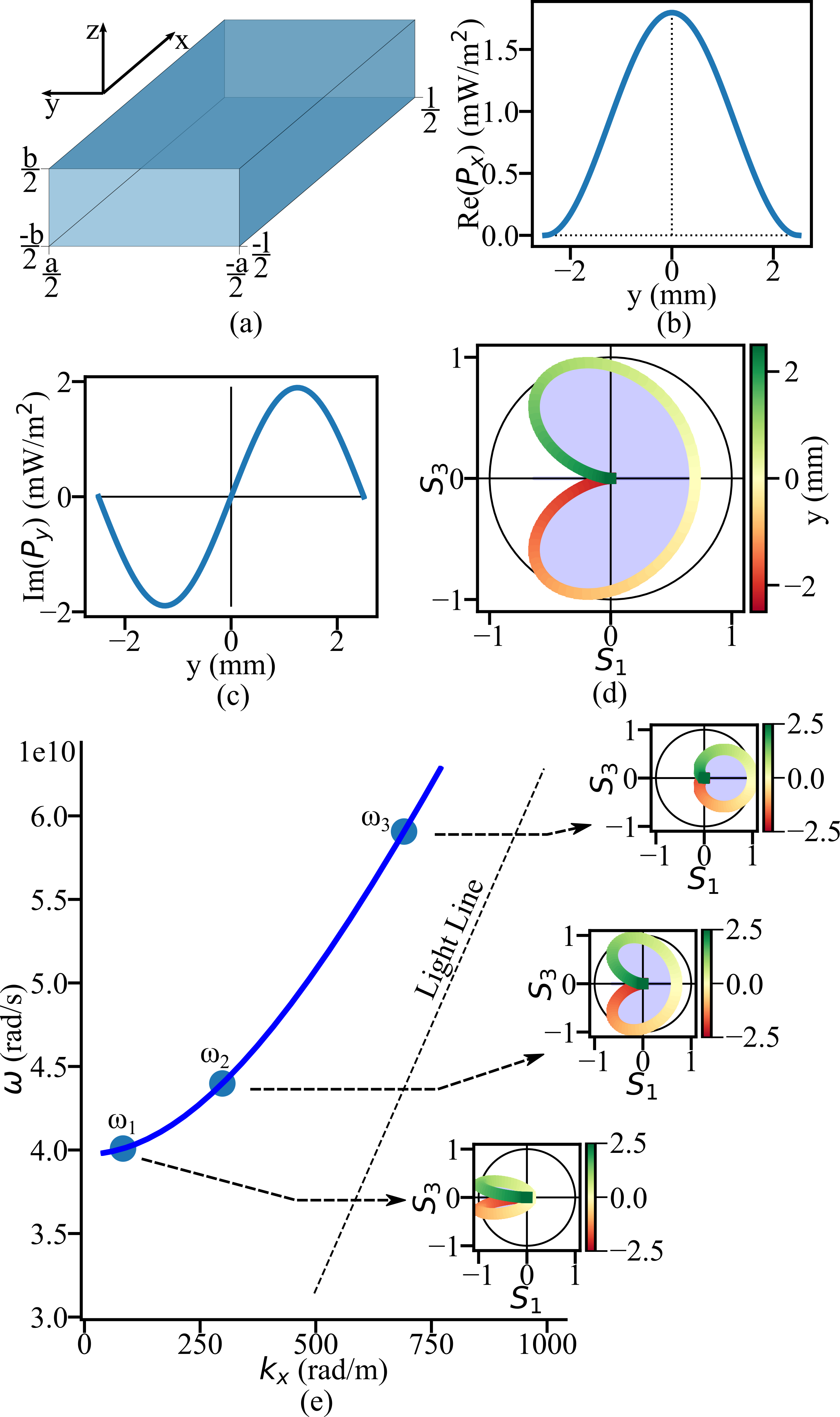}
    \caption{(a) Geometry of the dielectric-filled rectangular waveguide. (b) Longitudinal real and (c) transverse reactive components of the CTAPV. (d) Stokes plot over the Poincar\'e sphere for wave propagation across the dielectric-filled waveguide at $7$ GHz. (e) Dispersion curve corresponding to the dielectric-filled waveguide. Stokes plot representations at frequencies $\omega_1$ ($f_1=6.38$ GHz), $\omega_2$ ($f_2=7$ GHz) and $\omega_3$ ($f_3=9.4$ GHz). }
    \label{fig:diel_wg_analysis}
\end{figure}

We start our investigation with a benchmark scenario of a standard dielectric-filled non-gyrotropic rectangular waveguide of cross-sections $a=5$ mm and $b=3$ mm, as shown in Fig.~\ref{fig:diel_wg_analysis}(a). The parameters $a$, $b$ and $l$ correspond to the broad side dimension, narrow side dimension and length of the waveguide, respectively. The dielectric permittivity and permeability values are $\epsilon_r=15$ and $\mu_r=1.5$, respectively. Considering the fundamental $\text{TE}_\text{10}$ mode (the fundamental TE-to-x mode corresponding to guided-wave propagation in the $\hat{x}$-direction), we find the longitudinal real and transverse reactive power components of the complex CTAPV. The CTAPV has the form $\vec{P}_{\textrm{CTAPV}} = P_x\hat{x} + jP_y\hat{y}$ with
 \begin{equation}
 		P_x=\frac{0.5E_0^2k_x\cos^2(k_yy)}{\omega\mu_0},~
 		P_y=\frac{0.25E_0^2k_y\sin(2k_yy)}{\omega\mu_0}
 \end{equation}
 Here, the $k_y=\pi/a$ is the cutoff wave number and the dispersion relation is $k_x^2+k_y^2=\epsilon_r\epsilon_0\mu_r\mu_0\omega^2$. The plots corresponding to the $P_x$ and $P_y$ components across the dielectric waveguide at the frequency of $7$ GHz are shown in Fig.~\ref{fig:diel_wg_analysis}(b) and (c), respectively. We can observe the symmetric and anti-symmetric profiles of the real and reactive power components, respectively. The Poynting vector spin is in the $\hat{z}$ direction. Using these power components, we compute and show the Stokes plot in Fig.~\ref{fig:diel_wg_analysis}(d). $S_1$ and $S_3$ are normalized with respect to the maxima of $S_0$ along the cross-section. The location along the cross-section ($y$-axis) is color-mapped over the Stokes plot. 


 The Stokes plot representation for the dielectric-filled waveguide follows a cardioid like pattern which is symmetric across the $S_3=0$ line. The Stokes plot terminates at the origin of the Poincar\'e circle corresponding to $y=\pm a/2$, indicating that both real and reactive powers are zero at those points. The segments below and above this line correspond to the negative and positive spin of the IPV. The symmetric existence of the Stokes plot across the two sides of the of the $S_3=0$ line indicates equal and opposite spin of the IPV across the central axis, with $S_3=0$ along $y=0$.
 Further, the right and left region with respect to the $S_1=0$ line indicates the dominance of real power over reactive power and vice versa, respectively. If $|P_x|>|P_y|$ then $S_1$ is positive. In Fig.~\ref{fig:diel_wg_analysis}(e) we plot the Stokes plot at three different frequencies along the dispersion curve for the dielectric-filled waveguide, at $\omega_1$ ($f_1=6.38$ GHz), $\omega_2$ ($f_2=7$ GHz) and $\omega_3$ ($f_3=9.4$ GHz). We observe that the Stokes plot representation corresponding to $\omega_1$, which is near the cutoff frequency, has maximum occupancy on the left side of the $S_1=0$ line. Whereas, as we move away from the cutoff frequency, the Stokes plot shifts towards the right side of the $S_1=0$ line. Near the cutoff frequency, the reactive power is maximum, and the real power diminishes. On the other hand, at higher frequencies, the rightward shift of the Stokes plot indicates a relative increment of real power with respect to the imaginary power component along the cross-section. We can also correlate the lower group velocity near the cutoff frequency with increased net reactive power.
 
 Hence, the Stokes plot in this benchmark scenario conveys a comprehensive understanding of the nature of IPV spin and the relative magnitude of the real and reactive power components in the waveguide. With this reference, we will now investigate the additional effect of the gyrotropy on the power propagation in a ferrite-filled waveguide.

\subsection{Poynting vector spin in ferrite-filled waveguide}

\begin{figure}
    \centering
    \includegraphics[width=1\linewidth]{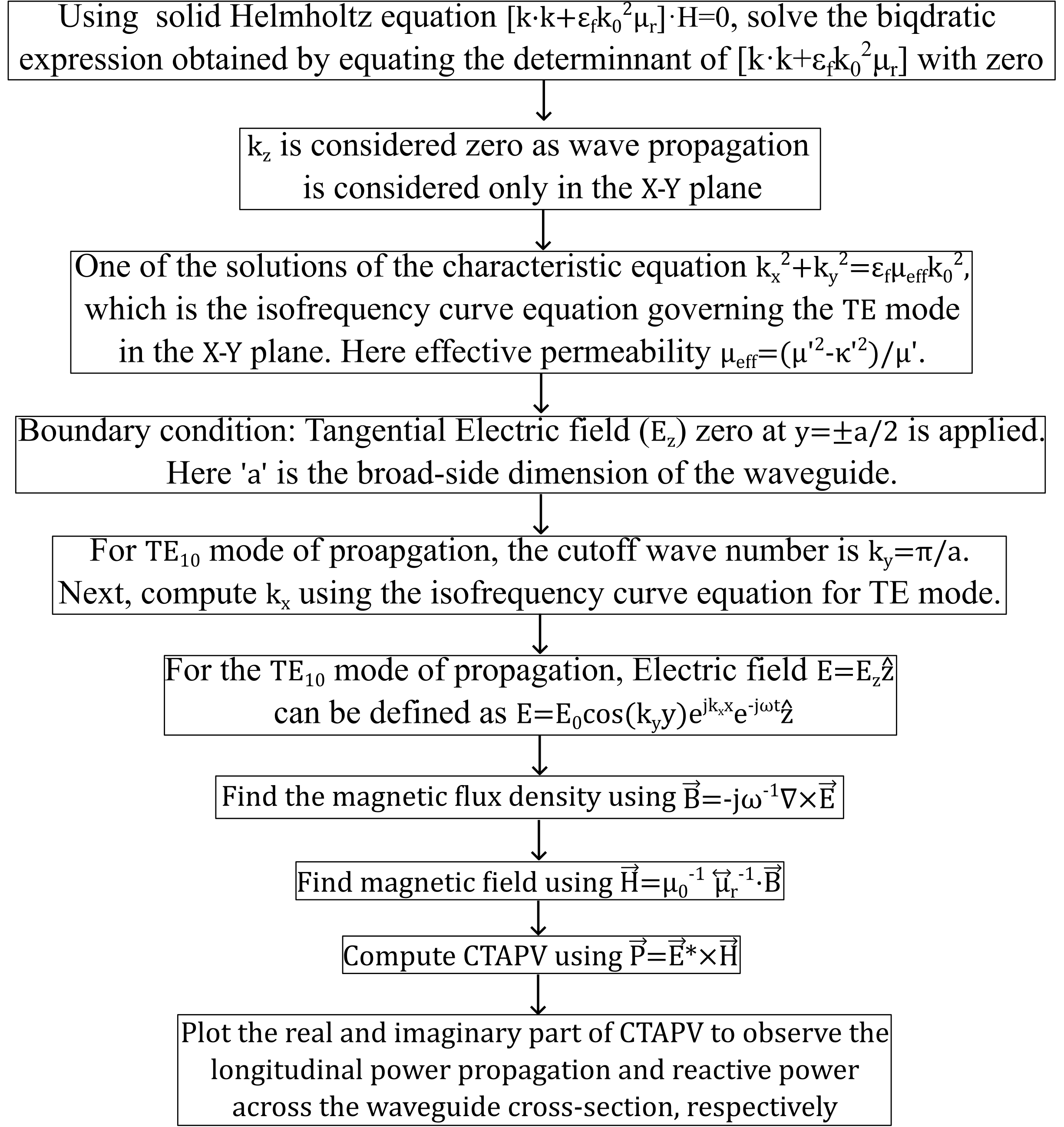}
    \caption{Flow chart describing the steps involved in the computation of fields and Poynting vector in a ferrite-filled waveguide, biased transverse to the direction of propagation.}
    \label{fig:flow_chart_2_waveguide}
\end{figure}

\begin{figure}
    \centering
    \includegraphics[width=\columnwidth]{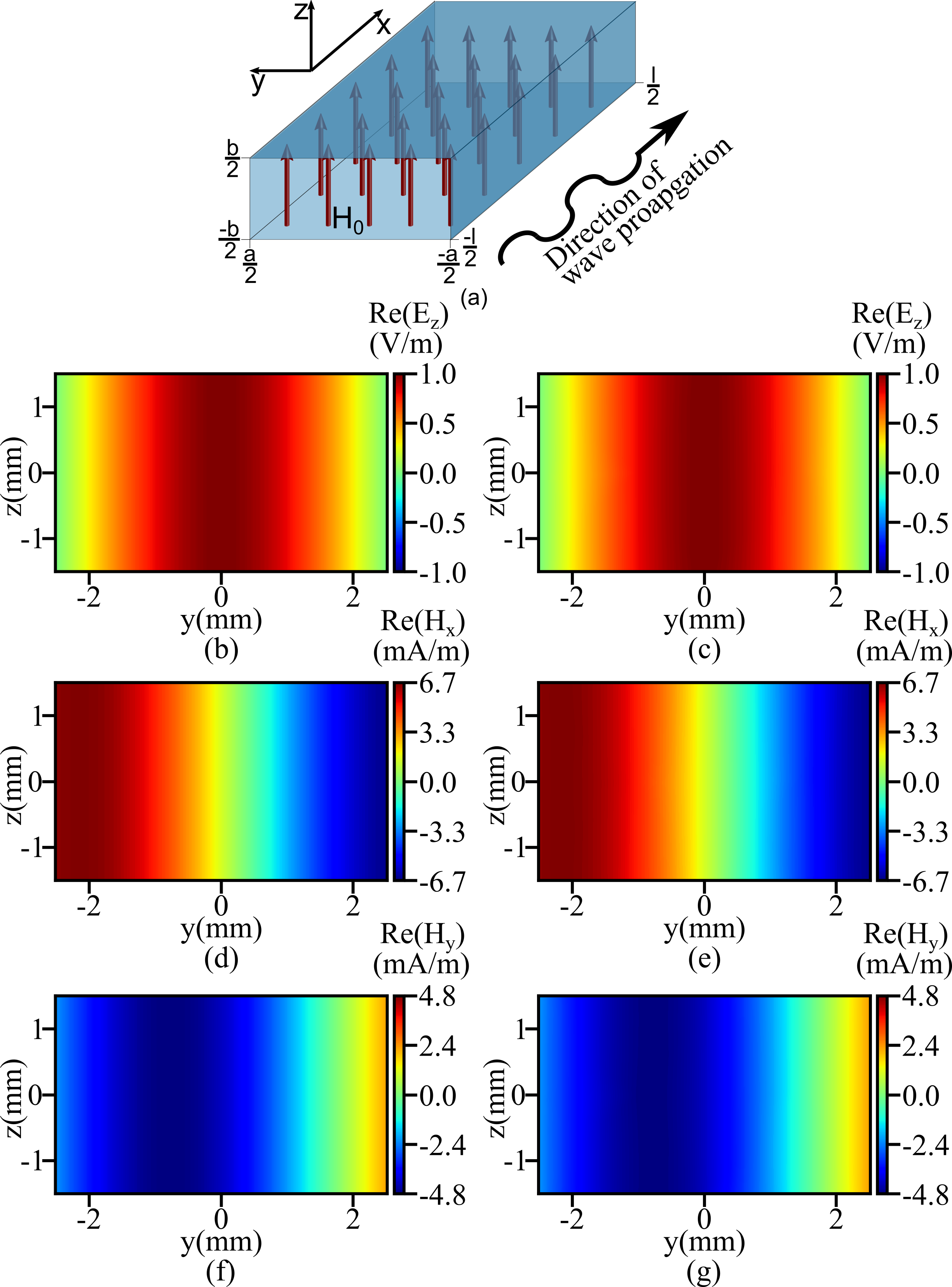}
    \caption{(a) Schematic of a $\hat{z}$-biased ferrite-filled rectangular waveguide. Field components along cross-section $Y-Z$ plane, (b,c) Re($E_z$) at $x=0$, (d,e) Re($H_x$) at $x=\lambda/4$ and (f,g) Re($H_y$) at $x=0$. Field computations in (b), (d), and (f) are done using~(\ref{eq:fer_wg_field_eqn}), whereas (c), (e), and (g) are obtained using full-wave simulation using CST Microwave Studio for validation.}
    \label{fig:ver_field_img}
\end{figure}

\begin{figure}
    \centering
    \includegraphics{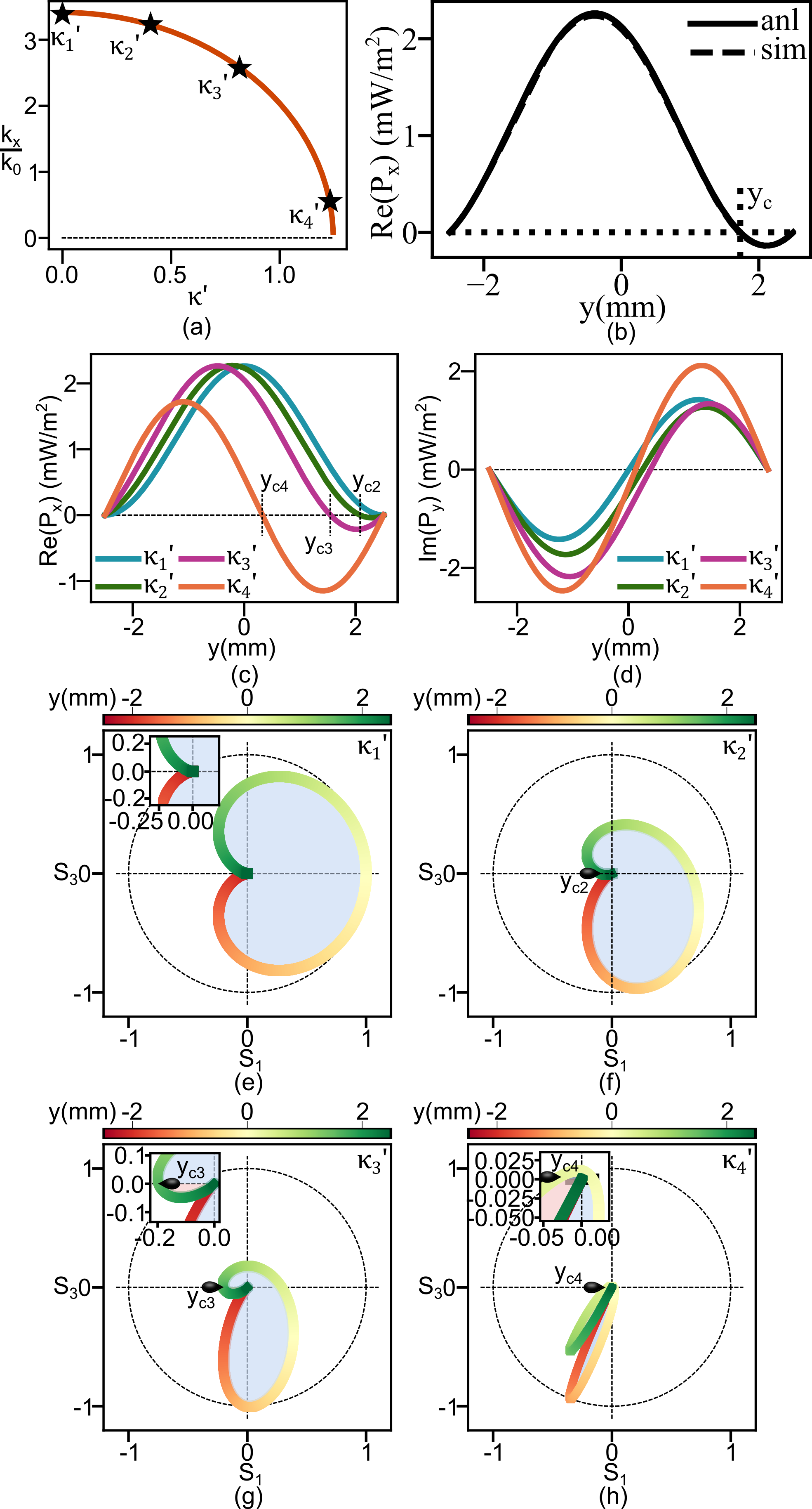}
    \caption{(a) variation of $k_x$ with respect to $\kappa^\prime$ variation from $0$ to the maximum value before $k_x$ reaches $0$. Four gyrotropy values $\kappa_1^\prime=0$, $\kappa_2^\prime=0.41$, $\kappa_3^\prime=0.82$, and $\kappa_4^\prime=1.23$ are marked. (b) Real component of Poynting vector Re$(Px)$ across the cross-section of the natural ferrite-filled waveguide computed at $f=7GHz$ and solved analytically and computed using full-wave simulation. The values of $\mu^\prime$ and $\kappa^\prime$ at this frequency are $1.98$ and $0.68$, respectively. (c) Longitudinal real and (d) transverse reactive components of CTAPV across the waveguide for the four gyrotropy values. (e)-(h) Stokes parameter plot across the cross-section of the waveguide for the four different gyrotropy values, respectively.}
    \label{fig:fer_wg_analysis}
\end{figure}

Fig.~\ref{fig:ver_field_img}(a) shows a rectangular waveguide filled with $\hat{z}$-biased YIG ferrite. The cross-section dimensions are selected as before ($a=5$ mm and $b=3$ mm). While again considering TE$_{10}$ mode, the cutoff wave-number comes out to be $k_y=\pi/a$. The steps for the computation of fields and Poynting vector are illustrated in Fig.~\ref{fig:flow_chart_2_waveguide}. The propagation constant for the fundamental TE mode of the ferrite-filled waveguide is,
\begin{equation}
	\label{eq:kx_fer_filled_wg_TE10_mode}
	k_x=\sqrt{\left(k_0^2\epsilon_f\frac{\mu^{\prime2}-\kappa^{\prime2}}{\mu^\prime}-k_y^2\right)}.
\end{equation} 
For a practical ferrite material both $\mu^\prime$ and $\kappa^\prime$ are functions of frequency. So the dispersion relation for an actual ferrite filled waveguide will be more complex. The dispersion relation and the corresponding stokes plots for practical YIG material are shown in the Supplementary material. The field equations for the TE$_\text{10}$ mode in the waveguide are (derived in the supplementary document)
\begin{equation}
	\label{eq:fer_wg_field_eqn}
	\begin{split}
		E_z&=E_0\cos(k_yy)e^{jk_xx}\\
		H_x&=j\frac{E_0(\mu^\prime k_y\sin(k_yy)-\kappa^\prime k_x\cos(k_yy))}{\omega\mu_0(\mu^{\prime2}-\kappa^{\prime2})}e^{jk_xx}\\
		H_y&=-\frac{E_0(\mu^\prime k_x\cos(k_yy)-\kappa^\prime k_y\sin(k_yy))}{\omega\mu_0(\mu^{\prime2}-\kappa^{\prime2})}e^{jk_xx}\\
	\end{split}
\end{equation}

We have to note here that the electric field component $E_z$ is identical to its non-gyrotropic form, whereas, gyrotropy effects the magnetic field components $H_x$ and $H_y$ by adding offset in the transverse direction because of $\kappa^\prime$. We consider the frequency of $7$ GHz, where the $\mu^\prime=1.98$ and $\kappa^\prime=0.68$ are both positive, corresponding to the elliptical regime. The analytically computed field components Re($E_z$), Re($H_x$) and Re($H_y$) across the waveguide cross-section $Y-Z$ plane using~\ref{eq:fer_wg_field_eqn} are shown in Fig.~\ref{fig:ver_field_img}(b), (d), and (f), at position $x=0$, $x=\lambda/4$ and $x=0$, respectively. Similarly, the field components computed using full-wave simulation using CST Microwave Studio is shown in Fig.~\ref{fig:ver_field_img}(c), (e), and (g), respectively, for validation. We can see a reasonable match between analytical and simulation results. In the case of ferrite-filled waveguide, while $E_z$ is symmetric corresponding to that of the usual TE$_{10}$ mode in a dielectric-filled waveguide, the magnetic field components are offset along the transverse direction. Especially, $H_y$ is important as it contributes to the real power propagation in the waveguide. We can see that $H_y$, in contrast to usual even and symmetric mode for conventional TE$_{10}$ mode, contains an additional region of opposite sign. This can be seen by the additional gyrotropy-induced red colored-region in addition to the expected blue-colored region. This is the region that will now support power propagation in the opposite direction, corresponding to wave propagation.

In order to shed more light on the effect of gyrotropy over wave propagation in the ferrite-filled waveguide, we select $f=7$ GHz and $\mu^\prime=2$ (which is the closest rounded value of $\mu^\prime=1.98$ at the chosen frequency). Next, we vary the value of gyrotropy of the medium by varying $\kappa^\prime$. With the help of~(\ref{eq:kx_fer_filled_wg_TE10_mode}) we can see that for fixed value of $k_y$, when we increase the magnitude of $\kappa^\prime$ the value of the $k_x$ decreases, and at some point before it reaches the value of $\mu^\prime$ (maximum possible value for TE mode propagation in bulk of the ferrite medium), $k_0\sqrt{\epsilon_f(\mu^{\prime2}-\kappa^{\prime2})/\mu^\prime}$ becomes equal to $k_y=\pi/a$ leading to $k_x=0$. Here, we get a gyrotropy-induced cutoff even when the medium supports the propagation of the bulk TE$_\text{10}$ mode. The variation of normalized $k_x/k_0$ with respect to gyrotropy term $\kappa^\prime$ is shown in Fig.~\ref{fig:fer_wg_analysis}(b). For our selected material parameters and waveguide dimensions, the gyrotropy-induced cutoff is attained at $\kappa^\prime=1.245$.

To develop a better understanding of this forward and backward power component, we analyze the expression for the CTAPV derived from~(\ref{eq:re_px_fer_wg}). The real and imaginary components of CTAPV has only $\hat{x}$- and $\hat{y}$-components, respectively, and are computed as Re$(\vec{P}_{CTAPV})\cdot\hat{x}=-0.5E_z^*H_y$ and Im$(\vec{P}_{CTAPV})\cdot\hat{y}=0.5E_z^*H_x$, leading to
\begin{eqnarray}
\label{eq:re_px_fer_wg}
    \text{Re}(P_x)=\frac{E_0^2\mu^\prime k_x\cos^2(k_yy)}{\omega\mu_0(\mu^{\prime2}-\kappa^{\prime2})}-\frac{E_0^2\kappa^\prime k_y\sin(k_yy)\cos(k_yy)}{\omega\mu_0(\mu^{\prime2}-\kappa^{\prime2})}\\
    \text{Im}(P_y)=\frac{E_0^2\mu^\prime k_y\sin(k_yy)\cos(k_yy)}{\omega\mu_0(\mu^{\prime2}-\kappa^{\prime2})}-\frac{E_0^2\kappa^\prime k_x\cos^2(k_yy)}{\omega\mu_0(\mu^{\prime2}-\kappa^{\prime2})}
\end{eqnarray}

We can see the positive and negative terms which govern the forward and backward power propagation with respect to the transverse $y$-coordinate. Note that, in the case of zero-gyrotropy ($\kappa^\prime=0$), the negative term in~(\ref{eq:re_px_fer_wg}) vanishes. It is the gyrotropic term $\kappa^\prime$ which induces this backward power propagation in such guiding structures. In Fig.~\ref{fig:fer_wg_analysis}(b), we plot Re$(P_x)$ corresponding to the natural values of ferrite at 7 GHz ($\mu^\prime=1.98$ and $\kappa^\prime=0.68$) using~(\ref{eq:re_px_fer_wg}) for analytical computation and CST Microwave Studio simulation. We can see the existence of the backward power flow in results and significant similarity between them. 

We can now find the exact point $y_c$ along $y$-axis which acts as the boundary between forward and backward power propagation. Using~(\ref{eq:re_px_fer_wg}), we equate Re$(P_x)=0$ and find $y_c=y|_{\text{Re}(P_x)=0}$ as
\begin{equation}
	\label{eq:y_c_pos_fer_filled_waveguide}
	y_c=\frac{a}{\pi}\tan^{-1}\left(\frac{\mu^\prime k_x}{\kappa^\prime k_y}\right),
\end{equation} 
This analytical~(\ref{eq:y_c_pos_fer_filled_waveguide}) may be important from the device engineering point of view to locate the null of real power propagation along the cross-section of the waveguide. This equation also shows that for no-gyrotropy conditions $\kappa^\prime=0$, $y_c$ shifts to $a/2$, which is at one of the walls of the waveguide. This position of $y_c$ at the wall for $\kappa^\prime=0$ means there is no backward power flow, which is consistent with power propagation in standard dielectric waveguides.

For further investigation, we select four specific values of $\kappa^\prime$. These gyrotropy points $\kappa_1^\prime=0$, $\kappa_2^\prime=0.41$, $\kappa_3^\prime=0.82$, and $\kappa_4^\prime=1.23$ are marked in Fig.~\ref{fig:fer_wg_analysis}(a). At the first point $\kappa_1^\prime=0$ the material acts as non-gyrotropic, whereas at $\kappa_4^\prime=1.23$ the gyrotropy is very near to the largest possible value before gyrotropy-induced cutoff is triggered in the waveguide. We can compute the real $P_x$ and imaginary $P_y$ of the CTAPV $\vec{P}_{avg}$ using the field equations given in~(\ref{eq:re_px_fer_wg}). Fig.~\ref{fig:fer_wg_analysis}(c) shows the longitudinal real $P_x$ power component for the different values of $\kappa^\prime$. We observe an important result: the gyrotropy leads to a decrease in the forward power propagation, and increase of an additional backward (negative) power flow region in the waveguide.

Note that for the three values $\kappa_1^\prime$, $\kappa_2^\prime$, and $\kappa_3^\prime$, we have marked three crossover points $y_{c2}$, $y_{c3}$, and $y_{c4}$, respectively, in Fig.~\ref{fig:fer_wg_analysis}(c). The transverse reactive power components Im$(P_y)$ for the four gyrotropy values are shown in Fig.~\ref{fig:fer_wg_analysis}(d). 
We can observe that the reactive power component maintains its sinusoidal variation at the extreme values of no gyrotropy ($\kappa_1^\prime=0$) and strong gyrotropy ($\kappa_1^\prime=1.23$). 
We use these real and reactive power components for the different $\kappa^\prime$ values to perform the Stokes plot analysis.

Fig.~\ref{fig:fer_wg_analysis}(e)-(h) shows the variation of the Stokes parameter with respect to the cross-section of the waveguide for gyrotropy value of $\kappa_1^\prime$, $\kappa_2^\prime$, $\kappa_3\prime$, and $\kappa_4^\prime$, respectively. For the first case of $\kappa_1^\prime=0$ corresponding to the non-gyrotropic case, the Stokes parameter follows a symmetrical cardioid like trace as was observed for the dielectric-filled waveguide. On the gradual increase of gyrotropy, we can see that most of the Stokes plot tends to shift to the lower half of the Poincar\'e circle. This primary occupancy of the lower half of Poincar\'e circle indicates the dominance of the negative $S_3$, which results from the negative spin of the IPV enforced by the gyrotropic material. It is as if the gyrotropic material is dragging the Stokes plot in the negative $S_3$ region.

Further, we observe that as a result of the increase in the gyrotropy, the Stokes parameter plot makes an additional transition across the $S_3=0$ line from the $S_3>0,S_1<0$ to the $S_3<0,S_1<0$ quadrant before finally terminating at the origin at the end positions $y=+a/2$. It is interesting to observe that, this transition across the $S_3=0$ axis takes place precisely at $y_c$, and the rest of the Stokes plot beyond the additional transition in the $S_3<0,S_1<0$ quadrant corresponds to the backward power propagation, thus showing the relation between the enhanced negative IPV spin region with the backward power propagation. This Stokes plot segment beyond $y=y_c$ up to $y=a/2$ increases in magnitude with a corresponding increase in $\kappa^\prime$, driving the cross-over point $y_c$ towards the center $y=0$ before the gyrotropy-induced cutoff is triggered. At the cutoff value of gyrotropy, both the lobes of the Stokes plot are divided in equal half by $y_c$. The existence of the Stokes parameter almost completely in the $S_1<0,S_3<0$ quadrant at the cutoff condition indicates the dominance of the negative spin of the IPV and the magnitude of reactive power component with respect to the real power propagation. At this point, we would like to highlight the similarity between the backward power flow observed here and that reported previously in surface waves \cite{2012Kim} and its dependence on the transverse reactive power flow, and rotational nature of IPV.

\begin{figure}
    \centering
    \includegraphics[width=\columnwidth]{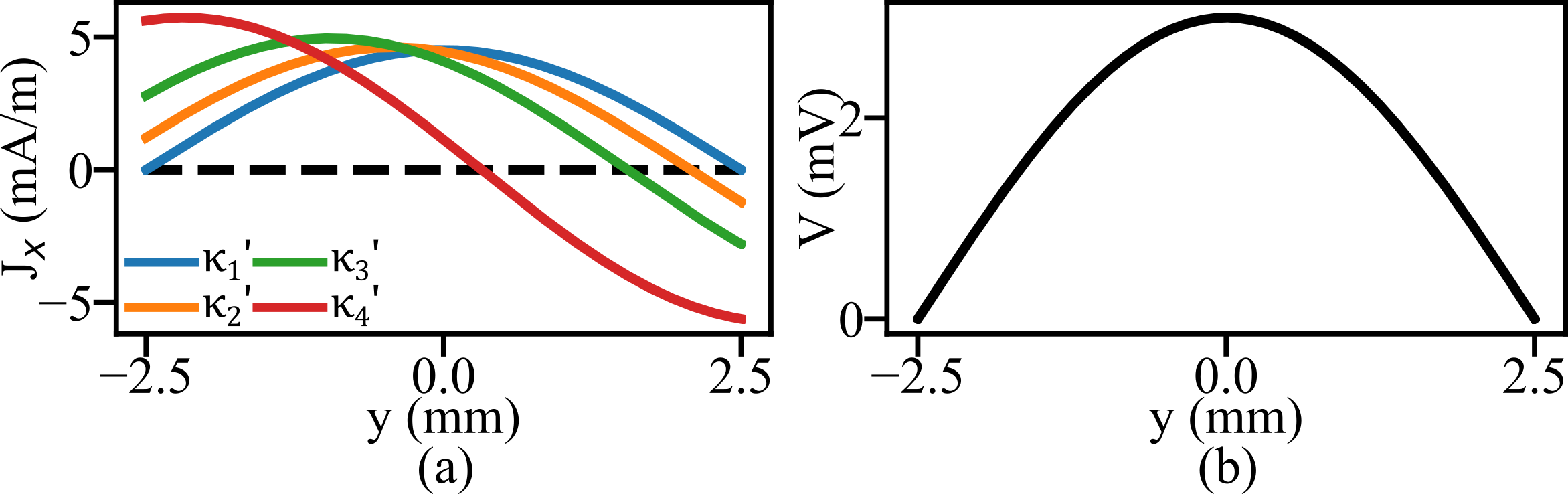}
    \caption{(a) Longitudinal component of surface current density $J_x$ at the top conducting plate ($z=b/2$) corresponding to gyrotropy values $\kappa_1^\prime=0$, $\kappa_2^\prime=0.41$, $\kappa_3^\prime=0.82$, and $\kappa_4^\prime=1.23$. Permeability term $\mu^\prime=2$ and frequency $f=7$ GHz. (b) Potential difference $V$ between the top and bottom conducting plates. $E_0$ is considered to be $1$ V/m.}
    \label{fig:long_surf_current}
\end{figure}

\subsection{Impact of negative power flow on surface currents}

The backward power flow in the ferrite-filled waveguide is also related to the surface current density on the top and bottom conducting walls of the waveguide. We compute the longitudinal component of surface current density at the top conducting plate ($z=+b/2$) as $J_x=(\hat{z}\times\vec{H})\cdot\hat{x}$ and plot it corresponding to four different gyrotropy values $\kappa_1^\prime=0$, $\kappa_2^\prime=0.41$, $\kappa_3^\prime=0.82$, and $\kappa_4^\prime=1.23$ in Fig.~\ref{fig:long_surf_current}(a). These computations were conducted considering $\mu^\prime=2$, and $f=7$ GHz. We can observe the oppositely directed part of $J_x$ with null points along $y_c$. In the supplementary files we have provided an analytically generated animation file showing the surface current distribution $\vec{J}$ over the top plate $y=+b/2$ of a rectangular waveguide, containing dielectric material in the first case and $\hat{z}$-biased ferrite with $\mu^\prime=2$ and $\kappa^\prime=0.6$ in the second case. The opposite direction of $J_x$ in the region $y_c\leq y\leq a/2$ and additional spin of surface current region stands in contrast with its dielectric-filled counterpart. Further, we provided the animation of $\vec{J}$ generated using full wave simulation using CST Microwave Studio for both top ($y=+b/2$) and bottom ($y=-b/2$) plates. The behavior of $\vec{J}$ observed at the plates matches with our analytically computed results. It is also worth noting that the location of the cross-over point remains the same for both the top and bottom plates, even though the direction of surface current propagation changes. With an increase in the gyrotropy of the medium, the opposite direction of $J_x$ increases.
The crossover point of the $J_x$ matches the crossover point $y_c$ of Re$(P_x)$ in Fig. \ref{fig:fer_wg_analysis}(c) for all the $\kappa^\prime$ values. Further, we can compute the potential difference between the top and bottom plates of the waveguides as $V=bE_z$. This potential difference $V$ with respect to the transverse $\hat{y}$-axis is shown in Fig.~\ref{fig:long_surf_current}(b). It can be seen that $V$ is independent of the effect of gyrotropy. This leads to an interesting scenario where the ratio between $V$ and $J_x$ is nonuniform along the $\hat{y}$-axis rather than being constant. Furthermore, this gyrotropy-induced variation of $J_x$ with respect to $V$ is consistent with the nonuniform wave impedance $Z_{wave}$ along the cross section of the waveguide. 
\begin{equation}
    Z_{wave}(y)=-\frac{E_z}{H_y}=\frac{\omega\mu_0(\mu^{\prime2}-\kappa^{\prime2})\cos(k_yy)}{k_x\mu^\prime\cos(k_yy)-\kappa^\prime k_y\sin(k_yy)}
\end{equation}
Taking these facts into consideration, we see that in the region where the real power propagates ($y_c<y<a/2$) in the backward direction, $Z_{wave}$ is negative and $J_x$ points in the opposite direction too. Specifically at the cross over point $y_c$, we have $Z_{wave}=\infty$ and $J_x=0$. When the direction of propagation reverses, $y_c$ also shifts to the other side of the waveguide, meaning the point with $Z_{wave}=\infty$ and $J_x=0$ is not same the for two counter-propagating waves. In general, we have nonreciprocal $Z_{wave}$ and $J_x$ across the cross-section. This may have important implications for the port terminal behavior of voltages and currents of such waveguides.

\section{ Stokes plots of surface wave along air-ferrite interface}
\label{sec:surf_wave_propagation}

\begin{figure}[!t]
    \centering
    \includegraphics[width=\columnwidth]{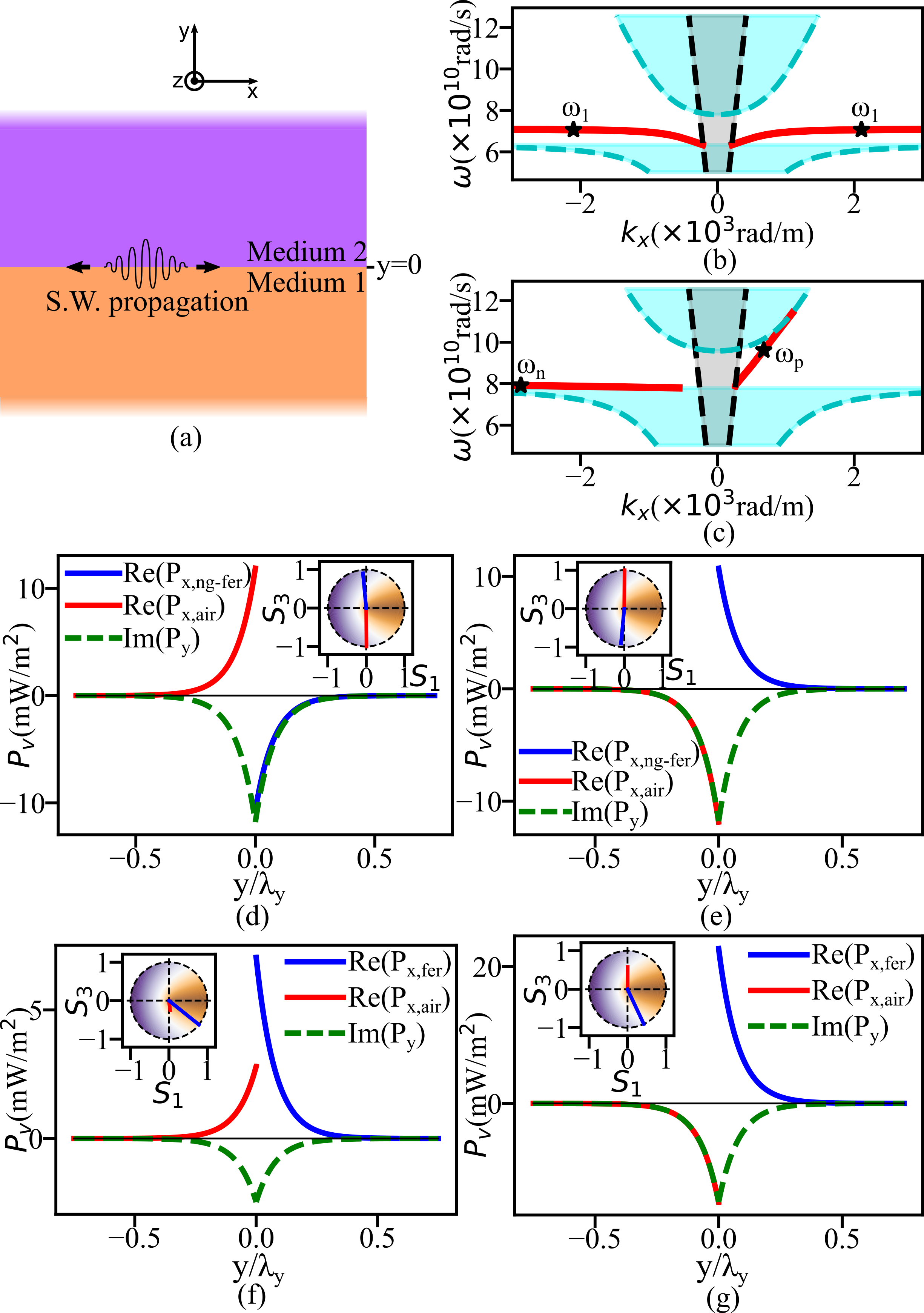}
    \caption{(a) Schematic representation of an interface between two media supporting SW propagation. Medium 1 is considered to be air and Medium 2 is considered to be NPM material in the first case and $-\hat{z}$-biased ferrite in the second case. (b) and (c) shows the dispersion plot corresponding to air-NPM and air-ferrite interfaces, respectively. Real and imaginary power components corresponding to air-NPM interface are shown in panel (d,e) and air-ferrite in (f,g). Further, (d) and (f) correspond to forward propagation, and (e) and (g) correspond to backward propagation. Stokes plot representation is added as insets to these plots with a colormap representation of the reactive power component.}
    \label{fig:air_ferrite_interface_analysis_new}
\end{figure}

Electromagnetic SW propagation along the interface between two media has been extensively investigated in literature \cite{Furs_2007,A_Hartstein_1973}. Since rotational nature of IPV and its link to backward power flow was first reported for surface waves at dielectric-negative-$\epsilon$ interfaces, it becomes important to conclude with the investigation of Poynting vector spin in dielectric-ferrite interface with negative $\mu$. Here, we look at the power flow and CTAPV associated with such propagation with the help of the Stokes plot. Fig.~\ref{fig:air_ferrite_interface_analysis_new}(a) represents an interface between two media, namely medium-1 and medium-2. These two mediums exist in the regions $y\leq0$ and $y\geq0$, respectively. We consider medium-1 to be air for simplicity.  To gain insight into the role of gyrotropy and IPV spin, we consider two cases. In the first case, medium-2 is a non-gyrotropic material (we will call it Negative Permeability Material, NPM) with $\kappa^\prime=0$ and frequency-dependent $\mu^\prime$, same as the ferrite material, which supports negative permeability in a specific frequency range. In the second case, medium-2 is considered as the $-\hat{z}$-biased ferrite material. The choice of these two media is done to highlight the difference between the transverse reactive power, spin of IPV, and oppositely directed real power propagation for non-gyrotropic and gyrotropic surface waves.

The material properties for the ferrite material are selected as given in Table~\ref{tab:mat_spec}. For both the cases, the dispersion relation and field equations can be found analytically and are presented in the supplementary document. The dispersion relation corresponding to the air-NPM and air-ferrite interfaces are shown in Fig.~\ref{fig:air_ferrite_interface_analysis_new}(b) and (c), respectively. The dispersion curves represented using red solid lines exhibit slow wave behavior. The gray and cyan shaded regions correspond to bulk propagation in air and ferrite media, respectively. In the first case of air-NPM interface, the surface waves are reciprocal. The real and imaginary components of CTAPV at a frequency of $\omega_1=2\pi\times11.25\times10^9$ rad/s are shown in Fig.~\ref{fig:air_ferrite_interface_analysis_new}(d) and (c) for the forward propagating and backward propagating mode, respectively. The frequency point is selected such that the group velocity is almost zero. At this frequency, the direction of real power flow along the interface is almost equal and opposite. However, it can be seen that the reactive power flow is continuous across the interface and is transverse to the direction of SW propagation. In \cite{2012Kim}, it was argued that the transverse reactive power accounts for the backward power flow. When the dispersion curve flattens out with zero group velocity, the magnitude of backward power flow in NPM is equal and opposite to the forward power flow. The spin of Poynting vector in these two regions are shown as inset Stokes plot. It can be seen that the IPV spin of propagating SW is opposite in the two media, and the sense of spin in both media reverses with the direction of propagation.

Fig.~\ref{fig:air_ferrite_interface_analysis_new}(f) and (g) shows the real and imaginary components of CTAPV for air-ferrite interface at frequencies $\omega_p=2\pi\times15.3\times10^9$ rad/s and $\omega_n=2\pi\times12.6\times10^9$ rad/s, corresponding to the forward and backward dispersion curve of the Fig.~\ref{fig:air_ferrite_interface_analysis_new} (c). It can be seen that, unlike the air-NPM interface, in this case, the real component of CTAPV in ferrite is positive for forward as well as backward SW propagation. The gyrotropy-induced spin in IPV ensures that the spin in ferrite is always negative. Therefore, the dispersion curve of the backward propagating mode flattens out - with zero net power flow - at a lower frequency, while the dispersion curve of the forward mode does not flatten out, and the net power flow and group velocity do not become zero. Therefore, understanding the combined effect of gyrotropy-induced IPV spin and the structure or geometry-induced spin is important for understanding gyrotropic SW propagation and its nonreciprocal nature.

\section{Implications for technological applications}
\label{sec:disc}
The primary focus of this paper is to understand and report the effect of gyrotropy on the reactive power flow and rotational nature of instantaneous Poynting vector waveguiding structures. However, the observation of a negative power flow in a region of cross-section is a non-trivial observation which can possibly have several technological implications.

The region of negative power flow and the fact that local average power flow has a null along the direction of wave propagation may result in region-dependent optical/radiation forces on particles. The rotational nature of the Poynting vector with the presence of transverse spin can have implications for guided waves with an additional degree of freedom, akin to orbital angular momentum or spatio-temporal modulation \cite{Zhang2025}. Since these properties can be controlled by an external magnetic field, it can support tunability as well.

The region of negative power flow has another interesting consequence. Because of gyrotropy, the current on the conducting plate of the waveguide loses its symmetry and has a region where the direction of the current is reversed. There is also a point along the cross-section of the plate where the current is zero. This is also expected to result in interesting transmission characteristics when such waveguides are interfaced with other microwave components, defects, and circuits.

Further, it should be noted the reported results are of more general significance because similar observations and analysis can be done for gyroelectric medium and magnetically biased plasma which are also regaining importance for several reasons. The presence of radiating structures in such medium with gyrotropy induced spin in fields and Poynting vector can also lead to interesting applications.

\section*{Acknowledgment}

We are thankful to the Department of Electronics and Communication Engineering, at North Eastern Regional Institute of Science and Technology, Arunachal Pradesh, India, for providing sufficient resources and support in this research.

\section{Conclusion}

\label{sec:conc}

We have analytically investigated the complex CTAPV in the bulk of gyromagnetic ferrite medium and showed the relation between the transverse reactive component and the spin of the IPV. We showed that even in bulk propagation in the gyromagnetic medium there is a reactive component on CTAPV, which causes spin in IPV. We quantified this spin and the relative strength of the power components using the method of Stokes plot representation. With the help of the Stokes plot we have shown the role of gyrotropy-governed reactive power over suppression of wave propagation. Further, we showed the asymmetrical nature of surface currents in such waveguides, and which is related to the nonuniform $Z_{wave}$ along the transverse direction of the waveguide. These results establish an important relationship between the spin of the Poynting vector and the backward power propagation in waveguiding structures involving gyrotropic materials. This understanding will be helpful in providing crucial design principles to realize novel tunable and nonreciprocal waveguiding applications. Our results corresponding to the nonuniform nature of current density and wave impedance in a ferrite-filled waveguide can be used to realize spatially varying terminal behavior for nonreciprocal applications.

\bibliographystyle{ieeetr}

\bibliography{reference.bib}

@article{2019Xu,
  title = {Azimuthal Imaginary Poynting Momentum Density},
  author = {Xu, Xiaohao and Nieto-Vesperinas, Manuel},
  journal = {Phys. Rev. Lett.},
  volume = {123},
  issue = {23},
  pages = {233902},
  numpages = {5},
  year = {2019},
  month = {Dec},
  publisher = {American Physical Society},
  doi = {10.1103/PhysRevLett.123.233902},
  url = {https://link.aps.org/doi/10.1103/PhysRevLett.123.233902}
}

@article{2012Kim,
  title = {Analysis of transverse power flow via surface modes in metamaterial waveguides},
  author = {Kim, Kyoung-Youm and Kim, Jungho and Lee, Il-Min and Lee, Byoungho},
  journal = {Phys. Rev. A},
  volume = {85},
  issue = {2},
  pages = {023840},
  numpages = {6},
  year = {2012},
  month = {Feb},
  publisher = {American Physical Society},
  doi = {10.1103/PhysRevA.85.023840},
  url = {https://link.aps.org/doi/10.1103/PhysRevA.85.023840}
}

@article{2022Sen,
  title = {Gyrotropy-governed isofrequency surfaces and photonic spin in gyromagnetic media},
  author = {Sen, Rajarshi and Pendharker, Sarang},
  journal = {Phys. Rev. A},
  volume = {105},
  issue = {2},
  pages = {023528},
  numpages = {10},
  year = {2022},
  month = {Feb},
  publisher = {American Physical Society},
  doi = {10.1103/PhysRevA.105.023528},
  url = {https://link.aps.org/doi/10.1103/PhysRevA.105.023528}
}

@book{Kong_2008, place={Cambridge, MA}, title={Electromagnetic wave theory}, publisher={EMW Publishing}, author={Kong, Jin Au}, year={2008}}

@article{2021Nieto,
  title = {Reactive helicity and reactive power in nanoscale optics: Evanescent waves. Kerker conditions. Optical theorems and reactive dichroism},
  author = {Nieto-Vesperinas, Manuel and Xu, Xiaohao},
  journal = {Phys. Rev. Res.},
  volume = {3},
  issue = {4},
  pages = {043080},
  numpages = {22},
  year = {2021},
  month = {Oct},
  publisher = {American Physical Society},
  doi = {10.1103/PhysRevResearch.3.043080},
  url = {https://link.aps.org/doi/10.1103/PhysRevResearch.3.043080}
}

@article{2012Litvin,
author = {Igor A. Litvin},
journal = {J. Opt. Soc. Am. A},
number = {6},
pages = {901--907},
publisher = {Optica Publishing Group},
title = {The behavior of the instantaneous Poynting vector of symmetrical laser beams},
volume = {29},
month = {Jun},
year = {2012},
url = {https://opg.optica.org/josaa/abstract.cfm?URI=josaa-29-6-901},
doi = {10.1364/JOSAA.29.000901},
}

@Article{Baumgartner2022,
AUTHOR = {Baumgartner, Paul and Masiero, Anna and Riener, Christian and Bauernfeind, Thomas},
TITLE = {Simulation Based Poynting Vector Description of the Field Regions for Simple Radiating Structures},
JOURNAL = {Electronics},
VOLUME = {11},
YEAR = {2022},
NUMBER = {13},
ARTICLE-NUMBER = {1967},
URL = {https://www.mdpi.com/2079-9292/11/13/1967},
ISSN = {2079-9292},
DOI = {10.3390/electronics11131967}
}

@Article{Picardi2019,
author={Picardi, Michela F.
and Neugebauer, Martin
and Eismann, J{\"o}rg S.
and Leuchs, Gerd
and Banzer, Peter
and Rodr{\'i}guez-Fortu{\~{n}}o, Francisco J.
and Zayats, Anatoly V.},
title={Experimental demonstration of linear and spinning Janus dipoles for polarisation- and wavelength-selective near-field coupling},
journal={Light: Science {\&} Applications},
year={2019},
month={Jun},
day={05},
volume={8},
number={1},
pages={52},
issn={2047-7538},
doi={10.1038/s41377-019-0162-x},
url={https://doi.org/10.1038/s41377-019-0162-x}
}

@ARTICLE{2019Afshani,
  author={Afshani, Amir and Wu, Ke},
  journal={IEEE Transactions on Microwave Theory and Techniques}, 
  title={Nonreciprocal Mode Converting Waveguide and Circulator}, 
  year={2019},
  volume={67},
  number={8},
  pages={3350-3360},
  doi={10.1109/TMTT.2019.2919591}}

@ARTICLE{2021Afshani,
  author={Afshani, Amir and Wu, Ke},
  journal={IEEE Transactions on Microwave Theory and Techniques}, 
  title={Generalized Theory of Concurrent Multimode Reciprocal and/or Nonreciprocal SIW Ferrite Devices}, 
  year={2021},
  volume={69},
  number={10},
  pages={4406-4421},
  doi={10.1109/TMTT.2021.3103990}}

@article{Fesenko2019,
  title = {Lossless and loss-induced topological transitions of isofrequency surfaces in a biaxial gyroelectromagnetic medium},
  author = {Fesenko, Volodymyr I. and Tuz, Vladimir R.},
  journal = {Phys. Rev. B},
  volume = {99},
  issue = {9},
  pages = {094404},
  numpages = {9},
  year = {2019},
  month = {Mar},
  publisher = {American Physical Society},
  doi = {10.1103/PhysRevB.99.094404},
  url = {https://link.aps.org/doi/10.1103/PhysRevB.99.094404}
}

@Article{Antognozzi2016,
author={Antognozzi, M.
and Bermingham, C. R.
and Harniman, R. L.
and Simpson, S.
and Senior, J.
and Hayward, R.
and Hoerber, H.
and Dennis, M. R.
and Bekshaev, A. Y.
and Bliokh, K. Y.
and Nori, F.},
title={Direct measurements of the extraordinary optical momentum and transverse spin-dependent force using a nano-cantilever},
journal={Nature Physics},
year={2016},
month={Aug},
day={01},
volume={12},
number={8},
pages={731-735},
issn={1745-2481},
doi={10.1038/nphys3732},
url={https://doi.org/10.1038/nphys3732}
}

@Article{Nieto-Vesperinas2022,
author={Nieto-Vesperinas, Manuel
and Xu, Xiaohao},
title={The complex Maxwell stress tensor theorem: The imaginary stress tensor and the reactive strength of orbital momentum. A novel scenery underlying electromagnetic optical forces},
journal={Light: Science {\&} Applications},
year={2022},
month={Oct},
day={12},
volume={11},
number={1},
pages={297},
issn={2047-7538},
doi={10.1038/s41377-022-00979-2},
url={https://doi.org/10.1038/s41377-022-00979-2}
}

@ARTICLE{Ghaffar2019,
  author={Ghaffar, Farhan A. and Bray, Joey R. and Vaseem, Mohammed and Roy, Langis and Shamim, Atif},
  journal={IEEE Transactions on Magnetics}, 
  title={Theory and Design of Tunable Full-Mode and Half-Mode Ferrite Waveguide Isolators}, 
  year={2019},
  volume={55},
  number={8},
  pages={1-8},
  doi={10.1109/TMAG.2019.2910028}}

@ARTICLE{Noferesti2021,
  author={Noferesti, Moein and Djerafi, Tarek},
  journal={IEEE Transactions on Magnetics}, 
  title={A Tunable Ferrite Isolator for 30 GHz Millimeter-Wave Applications}, 
  year={2021},
  volume={57},
  number={7},
  pages={1-7},
  doi={10.1109/TMAG.2021.3071684}}

@ARTICLE{Marzall2021,
  author={Marzall, Laila and Psychogiou, Dimitra and Popović, Zoya},
  journal={IEEE Transactions on Microwave Theory and Techniques}, 
  title={Microstrip Ferrite Circulator Design With Control of Magnetization Distribution}, 
  year={2021},
  volume={69},
  number={2},
  pages={1217-1226},
  doi={10.1109/TMTT.2020.3045995}}

@ARTICLE{Olivier2020,
  author={Olivier, Vincent and Huitema, Laure and Lenoir, Bertrand and Turki, Hamza and Breuil, Christophe and Pouliguen, Philippe and Monediere, Thierry},
  journal={IEEE Transactions on Microwave Theory and Techniques}, 
  title={Dual-Band Ferrite Circulators Operating on Weak Field Conditions: Design Methodology and Bandwidths’ Improvement}, 
  year={2020},
  volume={68},
  number={7},
  pages={2521-2530},
  doi={10.1109/TMTT.2020.2988003}}

@ARTICLE{Chou2018,
  author={Chou, Hsi-Tseng and Chang, Chia-Hung and Chen, Yen-Ting},
  journal={IEEE Transactions on Antennas and Propagation}, 
  title={Ferrite Circulator Integrated Phased-Array Antenna Module for Dual-Link Beamforming at Millimeter Frequencies}, 
  year={2018},
  volume={66},
  number={11},
  pages={5934-5942},
  doi={10.1109/TAP.2018.2862343}}

@ARTICLE{Nafe2015,
  author={Nafe, Ahmed and Shamim, Atif},
  journal={IEEE Transactions on Microwave Theory and Techniques}, 
  title={An Integrable SIW Phase Shifter in a Partially Magnetized Ferrite LTCC Package}, 
  year={2015},
  volume={63},
  number={7},
  pages={2264-2274},
  doi={10.1109/TMTT.2015.2436921}}

@ARTICLE{Kagita2017,
  author={Kagita, Srujana and Basu, Ananjan and Koul, Shiban K.},
  journal={IEEE Transactions on Magnetics}, 
  title={Characterization of LTCC-Based Ferrite Tape in ${X}$ -band and Its Application to Electrically Tunable Phase Shifter and Notch Filter}, 
  year={2017},
  volume={53},
  number={1},
  pages={1-8},
  doi={10.1109/TMAG.2016.2605078}}

@article{Xi_2021,
doi = {10.1088/1367-2630/ac1c84},
url = {https://dx.doi.org/10.1088/1367-2630/ac1c84},
year = {2021},
month = {aug},
publisher = {IOP Publishing},
volume = {23},
number = {8},
pages = {083042},
author = {Xiang Xi and Xi-Ming Li and Kang-Ping Ye and Hua-Bing Wu and Jian Chen and Rui-Xin Wu},
title = {Dual-polarization topological phases and phase transition in magnetic photonic crystalline insulator},
journal = {New Journal of Physics}
}

@Article{Zhou2020,
author={Zhou, Peiheng
and Liu, Gui-Geng
and Ren, Xin
and Yang, Yihao
and Xue, Haoran
and Bi, Lei
and Deng, Longjiang
and Chong, Yidong
and Zhang, Baile},
title={Photonic amorphous topological insulator},
journal={Light: Science {\&} Applications},
year={2020},
month={Jul},
day={24},
volume={9},
number={1},
pages={133},
issn={2047-7538},
doi={10.1038/s41377-020-00368-7},
url={https://doi.org/10.1038/s41377-020-00368-7}
}

@article{Tikhonov2019,
    author = {Tikhonov, V. V. and Litvinenko, A. N.},
    title = "{Spin-wave diagnostics of the magnetization distribution over the thickness of a ferrite film}",
    journal = {Applied Physics Letters},
    volume = {115},
    number = {7},
    pages = {072410},
    year = {2019},
    month = {08},
    issn = {0003-6951},
    doi = {10.1063/1.5098116},
    url = {https://doi.org/10.1063/1.5098116}}

@article{Tang2021,
    author = {Tang, Siyi and Sarker, Md Shamim and Ma, Kaijie and Yamahara, Hiroyasu and Tabata, Hitoshi and Seki, Munetoshi},
    title = "{Efficient spin-wave transmission in epitaxial thin films of defect spinel $\gamma$-$\text{Fe}_\text{2-x}$$\text{Al}_\text{x}$$\text{O}_\text{3}$}",
    journal = {Applied Physics Letters},
    volume = {119},
    number = {8},
    pages = {082402},
    year = {2021},
    month = {08},
    issn = {0003-6951},
    doi = {10.1063/5.0060102},
    url = {https://doi.org/10.1063/5.0060102}}

@article{Tuz2020,
    author = {Tuz, Vladimir R. and Fesenko, Volodymyr I.},
    title = "{Magnetically induced topological transitions of hyperbolic dispersion in biaxial gyrotropic media}",
    journal = {Journal of Applied Physics},
    volume = {128},
    number = {1},
    pages = {013107},
    year = {2020},
    month = {07},
    issn = {0021-8979},
    doi = {10.1063/5.0013546},
    url = {https://doi.org/10.1063/5.0013546},
    eprint = {https://pubs.aip.org/aip/jap/article-pdf/doi/10.1063/5.0013546/15247168/013107\_1\_online.pdf},
}

@article{Mukhopadhyay2022,
  title = {Anti-$\mathcal{PT}$ symmetry enhanced interconversion between microwave and optical fields},
  author = {Mukhopadhyay, Debsuvra and Nair, Jayakrishnan M. P. and Agarwal, Girish S.},
  journal = {Phys. Rev. B},
  volume = {105},
  issue = {6},
  pages = {064405},
  numpages = {8},
  year = {2022},
  month = {Feb},
  publisher = {American Physical Society},
  doi = {10.1103/PhysRevB.105.064405},
  url = {https://link.aps.org/doi/10.1103/PhysRevB.105.064405}
}

@article{Caloz2018,
  title = {Electromagnetic Nonreciprocity},
  author = {Caloz, Christophe and Al\`u, Andrea and Tretyakov, Sergei and Sounas, Dimitrios and Achouri, Karim and Deck-L\'eger, Zo\'e-Lise},
  journal = {Phys. Rev. Appl.},
  volume = {10},
  issue = {4},
  pages = {047001},
  numpages = {26},
  year = {2018},
  month = {Oct},
  publisher = {American Physical Society},
  doi = {10.1103/PhysRevApplied.10.047001},
  url = {https://link.aps.org/doi/10.1103/PhysRevApplied.10.047001}
}

@article{Abdelrahman2020,
  title = {Broadband and giant nonreciprocity at the subwavelength scale in magnetoplasmonic materials},
  author = {Abdelrahman, Mohamed Ismail and Monticone, Francesco},
  journal = {Phys. Rev. B},
  volume = {102},
  issue = {15},
  pages = {155420},
  numpages = {6},
  year = {2020},
  month = {Oct},
  publisher = {American Physical Society},
  doi = {10.1103/PhysRevB.102.155420},
  url = {https://link.aps.org/doi/10.1103/PhysRevB.102.155420}
}

@Book{dm_pozzar_book,
author={Pozar, David M.},
title={Microwave engineering},
year={2012},
publisher={Wiley; Fourth edition, India.},
}

@Book{erogolu_book,
author={Eroglu, Abdullah},
title={Wave Propagation and Radiation in Gyrotropic and Anisotropic Media},
year={2010},
publisher={Springer; New York.},
}

@ARTICLE{2024circulator1,
  author={Ali, Mohamed Mamdouh M. and Elsaadany, Mahmoud and Shams, Shoukry I. and Wu, Ke},
  journal={IEEE Transactions on Microwave Theory and Techniques}, 
  title={On the Design of Broadband Asymmetric Y-Junction Ferrite Circulator for 60-GHz Inter-Satellite Communication}, 
  year={2024},
  volume={72},
  number={2},
  pages={892-902},
  doi={10.1109/TMTT.2023.3298214}}

@ARTICLE{2023circulator2,
  author={Ashley, Andrea and Psychogiou, Dimitra},
  journal={IEEE Transactions on Microwave Theory and Techniques}, 
  title={Ferrite-Based Multiport Circulators With RF Co-Designed Bandpass Filtering Capabilities}, 
  year={2023},
  volume={71},
  number={6},
  pages={2594-2605},
  doi={10.1109/TMTT.2022.3233630}}

@ARTICLE{2023zouros,
  author={Zouros, Grigorios P. and Katsinos, Konstantinos},
  journal={IEEE Transactions on Antennas and Propagation}, 
  title={EM Scattering by Gyrotropic Circular Cylinders With Arbitrarily Oriented External Magnetic Bias}, 
  year={2023},
  volume={71},
  number={1},
  pages={1174-1179},
  doi={10.1109/TAP.2022.3218952}}

@ARTICLE{2022ueda,
  author={Ueda, Tetsuya and Kamino, Masaki and Kondo, Takumi and Itoh, Tatsuo},
  journal={IEEE Transactions on Microwave Theory and Techniques}, 
  title={Two-Degree-of-Freedom Control of Field Distribution on Nonreciprocal Metamaterial-Line Resonators and Its Applications to Polarization-Plane-Rotation and Beam- Scanning Leaky-Wave Antennas}, 
  year={2022},
  volume={70},
  number={1},
  pages={50-61},
  doi={10.1109/TMTT.2021.3124249}}

@ARTICLE{2022circulator3,
  author={Olivier, Vincent and Monediere, Thierry and Lenoir, Bertrand and Turki, Hamza and Breuil, Christophe and Pouliguen, Philippe and Huitema, Laure},
  journal={IEEE Transactions on Microwave Theory and Techniques}, 
  title={General Coupling Method for Stripline Ferrite Circulators: Application on Dual-Band Devices With Complex Central Conductor Shape}, 
  year={2022},
  volume={70},
  number={7},
  pages={3486-3497},
  doi={10.1109/TMTT.2022.3175168}}

@ARTICLE{2024ferritefilter,
  author={Yang, Mingyu and Wang, Haiyang and Ghannouchi, Fadhel M. and Li, Tianming and Zhou, Yihong and Li, Hao and Li, Qiwei and Shi, Liang and Ou, Meiling and Cheng, Renjie and Fu, Cheng and Liu, Xianqing and Hu, Biao},
  journal={IEEE Transactions on Microwave Theory and Techniques}, 
  title={Design of Filtering and Limiting Integrated Ferrite Device in High-Power Microwave Electromagnetic Environment}, 
  year={2024},
  volume={72},
  number={7},
  pages={3899-3907},
  doi={10.1109/TMTT.2023.3346905}}

@ARTICLE{2023Jemmeli,
  author={Jemmeli, S. and Monediere, Thierry and Arnaud, Eric and Huitema, Laure},
  journal={IEEE Transactions on Antennas and Propagation}, 
  title={Ultra-Miniature and Circularly Polarized Ferrite Patch Antenna}, 
  year={2023},
  volume={71},
  number={8},
  pages={6435-6443},
  doi={10.1109/TAP.2023.3284166}}

@ARTICLE{2023circulator4,
  author={Ali, Mohamed Mamdouh M. and Shams, Shoukry I. and Elsaadany, Mahmoud and Wu, Ke},
  journal={IEEE Transactions on Microwave Theory and Techniques}, 
  title={Stripline Y-Junction Circulators: Accurate Model and Electromagnetic Analysis Based on Gaussian Field Distribution Boundary Conditions}, 
  year={2023},
  volume={71},
  number={1},
  pages={146-155},
  doi={10.1109/TMTT.2022.3197161}}

@ARTICLE{2024GaoQ,
  author={Gao, Qian and Yeung, Lap K. and Liu, Yang and Geiler, Anton and Wang, Yuanxun Ethan},
  journal={IEEE Transactions on Microwave Theory and Techniques}, 
  title={Systematic Design of Planar Millimeter-Wave Filters Based on Hexagonal Barium Ferrite (BaM)}, 
  year={2024},
  volume={},
  number={},
  pages={1-13},
  doi={10.1109/TMTT.2024.3486309}}

@ARTICLE{2022Krowne,
  author={Krowne, Clifford M.},
  journal={IEEE Transactions on Microwave Theory and Techniques}, 
  title={Possibility of Raising the Frequency of RF Limiters Using the Nonlinear Spin Wave–Electromagnetic Interaction in Ferrites With High Saturation Magnetization}, 
  year={2022},
  volume={70},
  number={4},
  pages={2087-2097},
  doi={10.1109/TMTT.2021.3136230}}

@article{Shi_2022,
doi = {10.1088/2040-8986/ac5f22},
url = {https://dx.doi.org/10.1088/2040-8986/ac5f22},
year = {2022},
month = {apr},
publisher = {IOP Publishing},
volume = {24},
number = {5},
pages = {055103},
author = {Shi, Yuan-Kun and Wan, Bao-Fei and Zhang, Hai-Feng},
title = {Nonreciprocal Goos–Hänchen effect at the reflection of electromagnetic waves from the one-dimensional magnetized ferrite photonic crystals},
journal = {Journal of Optics}
}

@article{Yan:22,
author = {Hao Yan and Liqiao Jing and Jia Zhao and Chuanning Niu and Yujie Zhang and Liuge Du and Zuojia Wang},
journal = {Opt. Express},
number = {13},
pages = {24000--24008},
publisher = {Optica Publishing Group},
title = {Broadband nonreciprocal spoof plasmonic phase shifter based on transverse Faraday effects},
volume = {30},
month = {Jun},
year = {2022},
url = {https://opg.optica.org/oe/abstract.cfm?URI=oe-30-13-24000},
doi = {10.1364/OE.462863},
}

@article{Zhang_2024,
    author = {Zhang, Yujie and Jing, Liqiao and Niu, Chuanning and Zhao, Jia and Wang, Zuojia},
    title = {Broadband nonreciprocal gyromagnetic metasurface via magnetic Kerker-type dimers},
    journal = {Applied Physics Letters},
    volume = {124},
    number = {23},
    pages = {231701},
    year = {2024},
    month = {06},
    issn = {0003-6951},
    doi = {10.1063/5.0212581},
    url = {https://doi.org/10.1063/5.0212581},
    eprint = {https://pubs.aip.org/aip/apl/article-pdf/doi/10.1063/5.0212581/19973482/231701\_1\_5.0212581.pdf},
}

@ARTICLE{Deng2024,
  author={Deng, Bowen and Ding, Hao and Liu, Peiguo and Xu, Yanlin and Wu, Zhaofeng and Tian, Tao and Liu, Chenxi and Zha, Song},
  journal={IEEE Transactions on Microwave Theory and Techniques}, 
  title={Passive Adaptive Intensity High-Pass Devices for Nonreciprocal HPM Prevention}, 
  year={2024},
  volume={72},
  number={10},
  pages={5687-5698},
  doi={10.1109/TMTT.2024.3382297}}

@article{He:25,
author = {Jiafei He and Lin Zhang and Lingzhong Zhao and Hongxuan Mao and Qingtao Ba and Qilin Luo and Yiyun Chen and Huabing Wu and Shiyang Liu},
journal = {Opt. Lett.},
number = {1},
pages = {93--96},
publisher = {Optica Publishing Group},
title = {Lattice Kerker effect enabled single-layer nonreciprocal perfect absorbers by hybrid magnetic meta-atoms},
volume = {50},
month = {Jan},
year = {2025},
url = {https://opg.optica.org/ol/abstract.cfm?URI=ol-50-1-93},
doi = {10.1364/OL.546634},
}

@article{Grachev2024,
  title = {Nonreciprocal spin-wave transport in an asymmetric three-dimensional magnonic coupler},
  author = {Grachev, A.A. and Odintsov, S.A. and Beginin, E.N. and Sadovnikov, A.V.},
  journal = {Phys. Rev. Appl.},
  volume = {21},
  issue = {2},
  pages = {024031},
  numpages = {13},
  year = {2024},
  month = {Feb},
  publisher = {American Physical Society},
  doi = {10.1103/PhysRevApplied.21.024031},
  url = {https://link.aps.org/doi/10.1103/PhysRevApplied.21.024031}
}

@Article{Popov2023,
author={Popov, Maksym
and Xiong, Yuzan
and Zavislyak, Igor
and Chumak, Hryhorii
and Fedorchuk, Oleksandr
and Saha, Sujoy
and Bidthanapally, Rao
and Qu, Hongwei
and Page, Michael R.
and Srinivasan, Gopalan},
title={Y-type hexagonal ferrite-based band-pass filter with dual magnetic and electric field tunability},
journal={Scientific Reports},
year={2023},
month={Jan},
day={20},
volume={13},
number={1},
pages={1179},
issn={2045-2322},
doi={10.1038/s41598-023-28279-8},
url={https://doi.org/10.1038/s41598-023-28279-8}
}

@article{Chern:22,
author = {Ruey-Lin Chern and You-Zhong Yu},
journal = {Opt. Express},
number = {14},
pages = {25162--25176},
publisher = {Optica Publishing Group},
title = {Photonic topological semimetals in bigyrotropic metamaterials},
volume = {30},
month = {Jul},
year = {2022},
url = {https://opg.optica.org/oe/abstract.cfm?URI=oe-30-14-25162},
doi = {10.1364/OE.459097},
}

@ARTICLE{Zangeneh-Nejad2020,
  author={Zangeneh-Nejad, Farzad and Kaina, Nadège and Yves, Simon and Lemoult, Fabrice and Lerosey, Geoffroy and Fleury, Romain},
  journal={IEEE Transactions on Antennas and Propagation}, 
  title={Nonreciprocal Manipulation of Subwavelength Fields in Locally Resonant Metamaterial Crystals}, 
  year={2020},
  volume={68},
  number={3},
  pages={1726-1732},
  doi={10.1109/TAP.2019.2925927}}

@article{PhysRevB.109.085429,
  title = {Topological transitions and topological beam splitters in gyromagnetic metamaterials},
  author = {Li, Mingzhu and Han, Ning and Qi, Lu and Yan, Yu and Zhao, Rui and Liu, Shutian},
  journal = {Phys. Rev. B},
  volume = {109},
  issue = {8},
  pages = {085429},
  numpages = {10},
  year = {2024},
  month = {Feb},
  publisher = {American Physical Society},
  doi = {10.1103/PhysRevB.109.085429},
  url = {https://link.aps.org/doi/10.1103/PhysRevB.109.085429}
}

@article{R_Tuz_2024,
doi = {10.1088/1361-6463/ad19b6},
url = {https://dx.doi.org/10.1088/1361-6463/ad19b6},
year = {2024},
month = {jan},
publisher = {IOP Publishing},
volume = {57},
number = {13},
pages = {135005},
author = {R Tuz, Vladimir and Evlyukhin, Andrey B},
title = {Magneto-plasmonic scattering by a disk-shaped particle made of an artificial dielectric},
journal = {Journal of Physics D: Applied Physics}
}

@article{Halevi:81,
author = {P. Halevi and A. Mendoza-Hern\'{a}ndez},
journal = {J. Opt. Soc. Am.},
number = {10},
pages = {1238--1242},
publisher = {Optica Publishing Group},
title = {Temporal and spatial behavior of the Poynting vector in dissipative media: refraction from vacuum into a medium},
volume = {71},
month = {Oct},
year = {1981},
url = {https://opg.optica.org/abstract.cfm?URI=josa-71-10-1238},
doi = {10.1364/JOSA.71.001238},
}

@article{Mitri2016,
  title = {Superposition of nonparaxial vectorial complex-source spherically focused beams: Axial Poynting singularity and reverse propagation},
  author = {Mitri, F. G.},
  journal = {Phys. Rev. A},
  volume = {94},
  issue = {2},
  pages = {023801},
  numpages = {10},
  year = {2016},
  month = {Aug},
  publisher = {American Physical Society},
  doi = {10.1103/PhysRevA.94.023801},
  url = {https://link.aps.org/doi/10.1103/PhysRevA.94.023801}
}

@book{JD_jackson_book,
  address = {New York, {NY}},
  author = {Jackson, John David},
  edition = {3rd ed.},
  isbn = {9780471309321},
  lccn = {538.3537.8},
  publisher = {Wiley},
  title = {Classical electrodynamics},
  url = {http://cdsweb.cern.ch/record/490457},
  year = 1999
}

@Article{Zhang2025,
author={Zhang, Ruo-Yang
and Cui, Xiaohan
and Zeng, Yuan-Song
and Chen, Jin
and Liu, Wenzhe
and Wang, Mudi
and Wang, Dongyang
and Zhang, Zhao-Qing
and Wang, Neng
and Wu, Geng-Bo
and Chan, C. T.},
title={Bulk--spatiotemporal vortex correspondence in gyromagnetic zero-index media},
journal={Nature},
year={2025},
month={May},
day={01},
volume={641},
number={8065},
pages={1142-1148},
issn={1476-4687},
doi={10.1038/s41586-025-08948-6},
url={https://doi.org/10.1038/s41586-025-08948-6}
}

@article{Furs_2007,
doi = {10.1088/1751-8113/40/2/010},
url = {https://dx.doi.org/10.1088/1751-8113/40/2/010},
year = {2006},
month = {dec},
publisher = {},
volume = {40},
number = {2},
pages = {309},
author = {Furs, A N and Barkovsky, L M},
title = {A new type of surface polaritons at the interface of the magnetic gyrotropic media},
journal = {Journal of Physics A: Mathematical and Theoretical}
}

@article{A_Hartstein_1973,
doi = {10.1088/0022-3719/6/7/016},
url = {https://dx.doi.org/10.1088/0022-3719/6/7/016},
year = {1973},
month = {apr},
publisher = {},
volume = {6},
number = {7},
pages = {1266},
author = {A Hartstein and E Burstein and A A Maradudin and R Brewer and R F Wallis},
title = {Surface polaritons on semi-infinite gyromagnetic media},
journal = {Journal of Physics C: Solid State Physics}
}

@article{2025Sen,
  title = {Photonic-spin-governed absorption in a $\ensuremath{\mu}$-near-zero subwavelength gyromagnetic cylinder},
  author = {Sen, Rajarshi and Pendharker, Sarang},
  journal = {Phys. Rev. A},
  volume = {112},
  issue = {2},
  pages = {023506},
  numpages = {10},
  year = {2025},
  month = {Aug},
  publisher = {American Physical Society},
  doi = {10.1103/8kyf-gcd1},
  url = {https://link.aps.org/doi/10.1103/8kyf-gcd1}
}

\end{document}